\documentclass[amsmath,eqsecnum,preprint,pre,aps,showpacs]{revtex4}
\usepackage{amsmath}
\usepackage{graphicx}
\usepackage{bm}
\usepackage{float}

\begin{document}

\draft

\title{Dynamic equivalences in the Hard-sphere Dynamic Universality
Class}

\author{
Leticia L\'opez-Flores$^1$,
Honorina Ru\'iz-Estrada$^1$,
Mart\'in Ch\'avez-P\'aez$^2$,
Magdaleno Medina-Noyola$^2$
}

\address{
$^1$Facultad de Ciencias Fisico-Matem\'aticas,
Benem\'{e}rita Universidad Aut\'{o}noma de Puebla,
Apartado Postal 1152,
72000 Puebla, Pue.,
M\'{e}xico}

\address{$^{2}$Instituto de F\'{\i}sica {\sl ``Manuel Sandoval Vallarta"},
Universidad Aut\'{o}noma de San Luis Potos\'{\i}, \'{A}lvaro
Obreg\'{o}n 64, 78000 San Luis Potos\'{\i}, SLP, M\'{e}xico}

\date{\today}

\begin{abstract}

We perform systematic simulation experiments on model systems with
soft-sphere repulsive interactions to test the predicted dynamic
equivalence between soft-sphere liquids with similar static
structure. For this we compare the simulated dynamics (mean
squared displacement, intermediate scattering function,
$\alpha$-relaxation time, etc.) of different soft-sphere systems,
between them and with the hard-sphere liquid. We then show that
the referred dynamic equivalence does not depend on the (Newtonian
or Brownian) nature of the microscopic laws of motion of the
constituent particles, and hence, applies independently to
colloidal and to atomic simple liquids. Finally, we verify another
more recently-proposed dynamic equivalence, this time between the
\emph{long-time} dynamics of an atomic liquid and its
corresponding Brownian fluid (i.e., the Brownian system with the
same interaction potential).
\end{abstract}

\pacs{23.23.+x, 56.65.Dy}
\maketitle

\section{Introduction}\label{sec1}

At first sight, the macroscopic dynamics of supercooled liquids
seems to be strongly  material-specific, with no universal
character at all. This is evidenced, for example, by the great
diversity of molecular glass formers (ionic, metallic, organic,
polymeric, etc.), giving rise to an overwhelmingly rich
phenomenology
\cite{ngaireview1,angellreview1,edigerreview1,angell,debenedetti}.
One can easily understand this lack of universal behavior in
terms of the wide differences in the materials' structure and
composition, the masses of their individual atoms, their
preparation protocol, etc. Of course, the scenario becomes
even more complex when one attempts to include colloidal
systems in the discussion.

The formation of colloidal glasses and gels has been the subject
of intense study during the last two decades \cite{sciortinotartaglia},
and it is a widespread notion that the phenomenology of both,
the glass transition in ``thermally-driven" molecular glass formers,
and the dynamic arrest transition in ``density-driven" hard-sphere
colloidal systems, might share a common underlying universal
origin \cite{jammingrheology}. Two relevant conceptual issues,
however, must be understood in order for this expectation to
have a more fundamental basis. The first one requires us to spell
out the manner in which undercooling an atomic liquid might be
equivalent to overcompressing a colloidal liquid. The second
is to clarify under what conditions the macroscopic dynamics
of both classes of systems could be expected to be equivalent,
given the fact that the microscopic dynamics is Newtonian in
atomic liquids and Brownian in colloidal fluids.

The answer to these two questions is highly relevant since
it will allow us to understand which aspects of the macroscopic
dynamics of a given system are universal and which ones are
system-specific. These two issues have been addressed using
computer simulation methods on well defined model systems.
For example, interesting scalings of the \emph{equilibrium}
dynamics of simple models of soft-sphere glass formers have been
exposed by systematic computer simulations \cite{xu,berthierwitten1},
which provide an initial clue to the possible physical origin
of the equivalence between the process of cooling and the
process of compression. Similarly, also using computer simulations,
it has been partially corroborated that standard molecular dynamics
will lead to essentially the same dynamic arrest scenario as
Brownian dynamics  for a given model system (i.e., same
pair potential) \cite{lowenhansenroux,szamelflenner,puertasaging}.

From the theoretical side it would be desirable to have a unified
description of the macroscopic dynamics of both, colloidal and
atomic liquids, which explicitly predicts the aspects of the
macroscopic dynamics that are expected to be universal. These
topics might be addressed in the framework of a theory such as the
mode coupling theory of the ideal glass transition \cite{goetze1}.
In fact, the similarity of the long-time dynamics of Newtonian and
Brownian systems in the neighborhood of the glass transition, for
example, has been studied within this theoretical framework
\cite{szamellowen}. A number of issues, however, still remain open
\cite{szamelflenner,dyrejpcm2013}.

The present paper is part of an effort aimed at addressing these
two fundamental issues within a general  theoretical framework,
namely, the generalized Langevin equation (GLE)
formalism \cite{boonyip,delrio,faraday}. This formalism was employed
in the construction of the self-consistent generalized Langevin
equation (SCGLE) theory of colloid
dynamics \cite{scgle0,scgle1,scgle2}, eventually applied to the
description of dynamic arrest phenomena \cite{rmf,todos1,todos2},
and more recently, to the construction of a first-principles
theory of equilibration and aging of colloidal glass-forming
liquids \cite{noneqscgle0,noneqscgle1}.

When applied to model systems with soft repulsive
interactions  \cite{soft1}, the SCGLE theory of colloid dynamics,
together with the condition of static structural equivalence
between soft- and hard-sphere systems, predicts the existence
of a ``hard-sphere dynamic universality class", constituted
by the soft-sphere systems whose dynamic parameters, such as
the $\alpha$-relaxation time and self-diffusion coefficient,
depend on density, temperature and softness in a  universal
scaling fashion  \cite{soft2}, through an effective hard-sphere
diameter determined by the  Andersen-Weeks-Chandler \cite{awc,hansen}
criterion. These predictions provide a more fundamental explanation
of the scalings previously exhibited by computer
simulations \cite{xu,berthierwitten1}, and point to the physical
basis of the dynamic equivalence between cooling and compressing.

The main purpose of this paper is to report the results of,
and to provide detailed technical information on, a number of
simulation experiments performed with the purpose of testing
this density-temperature-softness scaling in the referred dynamic
universality class. An illustrative selection of these results
were advanced in a recent brief communication \cite{soft2}.
The second main purpose of the present paper is to perform the
pertinent simulation experiments to test a second relevant
prediction of the SCGLE theory, which addresses the second
of the two fundamental issues mentioned above, namely, the
\textit{macroscopic} dynamic equivalence between atomic and colloidal
liquids. As it happens, the SCGLE theory of colloid dynamics
is being extended to describe the dynamics of simple atomic
liquids \cite{atomic1,atomic2}. The scenario that emerges from
these theoretical developments include well defined scaling rules
that exhibit the equivalence between the dynamics of colloidal
fluids and the \emph{long-time} dynamics of atomic liquids. Here
we test these scalings by comparing the simulation results
for a given model system using both, molecular dynamics and
Brownian dynamics simulations.

Thus, the present paper is essentially a report of a set of
systematic computer simulations. In Sec. \ref{sec2} we define the
model systems considered in our study and provide the basic
information on the simulation methods  employed. In Sec.
\ref{sec3} we review the concept of static structural equivalence
between soft- and hard-sphere fluids, and explain how this concept
is employed to map the static structure of any soft-sphere liquid
onto the properties of an effective hard-sphere liquid. In Sec.
\ref{sec4} we review the extension of this structural equivalence
to the dynamic domain and present the simulation results that
validate the accuracy of the resulting  dynamic equivalence
between soft- and hard-sphere liquids. We first verify that this
dynamic equivalence is exhibited by our Brownian dynamics
simulations, and then confirm that the same dynamic equivalence is
also observed in the results of our molecular dynamics
simulations. In Sec.  \ref{sec5} we explain the correspondence
between the dynamics of colloidal fluids and the \emph{long-time}
dynamics of atomic liquids, and verify that the predicted scalings
are indeed satisfied by our molecular and Brownian dynamics
simulations. In the last section, Sec. \ref{sec6},  besides
summarizing the main results of this paper, we explain that for
the systems with interaction potential in the hard sphere dynamic
universality class, these effects can be taken into account
through the value of the short-time self-diffusion coefficient,
thus expanding the applicability of the scalings discussed here.
At this point we have to mention that the present study only
involves Brownian dynamics simulations that completely ignore the
effects of hydrodynamic interactions, which have an enormous
practical relevance in concentrated colloidal fluids.

\section{Methodological aspects} \label{sec2}

In this section we describe the most relevant methodological
aspects of this work. This includes information on the
numerical simulation methods and on the theoretical concepts
and approaches employed.

\subsection{Model potentials}\label{subsec2.a}
Let us consider a model liquid formed by \emph{N} spherical
particles in a volume \emph{V} which interact through a soft
repulsive pair potential $u(r)$ with tunable softness. We intend
to study the interplay of the effects of the number density (or
concentration, in the case of colloidal liquids) $n\equiv N/V$,
temperature $T$, and softness, represented by some parameter
denoted generically as $\nu$. There is a variety of analytic
proposals for such tunable soft potential
\cite{xu,berthierwitten1}, but for concreteness here we shall
refer explicitly to three specific representative model systems.
The first is the truncated Lennard-Jones (TLJ) potential,

\begin{equation}
u^{(\nu)}(r)=\epsilon\left[
\left( \frac{\sigma}{r} \right)^{2\nu}
-2 \left( \frac{\sigma}{r}\right)^{\nu} + 1
\right]\Theta(\sigma-r)   ,
\label{lj.pot}
\end{equation}

\noindent
in which $\Theta(x)$ is the unit step function.
The positive parameter $\nu$ controls the softness of the
interaction, with the limit $\nu \to \infty$ corresponding
to the hard sphere potential between particles of diameter
$\sigma$. For fixed $\nu$, the state space of this system is
spanned by the dimensionless temperature $T^*\equiv k_BT/\epsilon$
and volume fraction $\phi = \pi n \sigma^3/6$.

The second representative model system we shall refer to is the
inverse power-law (IPL) potential
\begin{equation}
u(r)=\epsilon (\sigma/r)^{2\nu},
\label{ipl.pot}
\end{equation}

\noindent
commonly used to model hard sphere effects. Just like in the
case of the TLJ model, for fixed softness parameter  $\nu$
the state space of the IPL model system is also spanned by
the dimensionless temperature $T^*\equiv k_BT/\epsilon$ and
volume fraction $\phi = \pi n \sigma^3/6$. The fundamental
difference between the IPL potential and the TLJ interaction
is that the latter is always short-ranged.

The third interaction model that we shall refer to is defined
by the hard-sphere plus repulsive Yukawa (HSY) potential,
frequently used to model the screened electrostatic repulsions
between charged colloidal particles \cite{nagele0}.
This is defined here as

\begin{equation}
u(r)=\epsilon \left[
\frac{ \exp [-z(r/\sigma-1)] } {(r/\sigma) }
\right]   .
\label{yuk.pot}
\end{equation}

\begin{figure}[htmb]
\centering
\includegraphics[width=7.0cm]{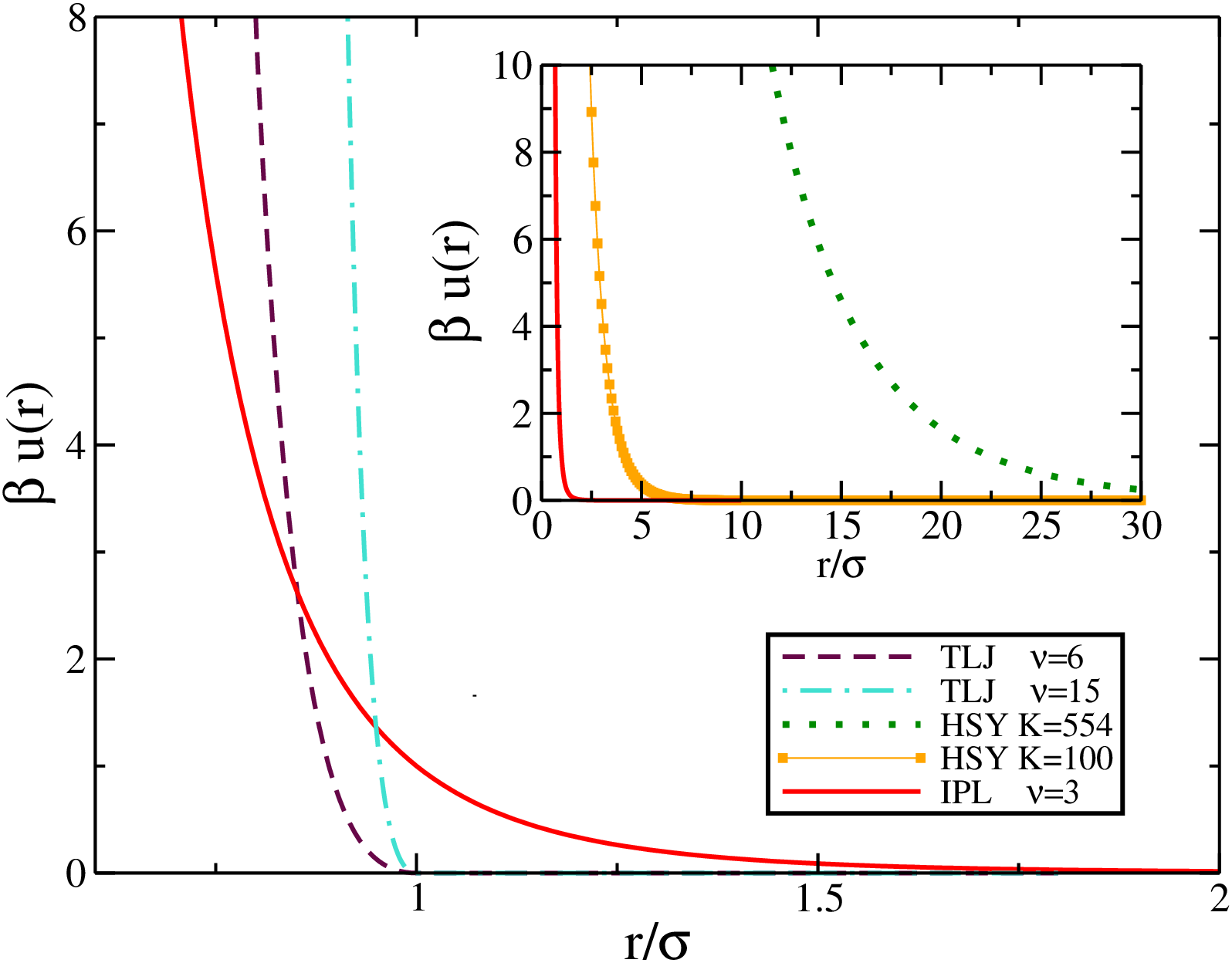}
\caption{ Illustration of the truncated Lennard-Jones (TLJ),
inverse power-law (IPL), with $T^{*}=1$ and  hard-sphere plus repulsive Yukawa
(HSY) potentials, with $K=100$ and $K=554$.} \label{potes.fig}
\end{figure}

\noindent
For fixed screening parameter $z$, the state space of this
system is also spanned by the volume fraction
$\phi = \pi n \sigma^3/6$ and the dimensionless temperature
$T^*\equiv k_BT/\epsilon$ (sometimes, however, we shall also
refer to the repulsion intensity parameter
$K\equiv 1/T^*=\beta\epsilon$).  The inverse screening
length $z$ controls the range of the potential, and for
our purpose, we may consider that it plays the role of the
softness parameter. Typical values for these parameters
representing real suspensions of highly charged colloidal
suspensions at low ionic strength in the dilute regime
are $K=554$, $z=0.149$, and $\phi$ of the order
of $10^{-4}$  \cite{gaylor}. We shall use these as illustrative
values, along with $K=100$ and $z=1.0$. Figure \ref{potes.fig}
plots these interaction models for some specific values of
these parameters to illustrate the variety of interactions considered.

\subsection{Simulations}\label{subsec2.b}
Molecular dynamics (MD) simulations using the velocity-verlet
algorithm \cite{allen} were conducted for the model liquids above,
formed by \emph{N} spherical particles of mass $M$ in a simulation
box of volume \emph{V}. The results are expressed in the well
known Lennard-Jones units, where $M$, $\sigma$ and  $\epsilon$ are
taken as the units of mass, length, and energy, respectively, and
$t_{MD}=\sqrt{M\sigma^2/k_B T}$ is the corresponding time unit.

For Brownian dynamics (BD) simulations we follow the prescription
proposed by Ermak and McCammon \cite{allen} to evolve the
positions of the particles in the simulation box. Thus, a given
particle at position $\vec r(t)$ and under the force $\vec F(t)$
is displaced in the $\alpha$ direction according to

\begin{equation}
r_{\alpha}(t+\Delta t)= r_{\alpha}(t) +
\beta D^0 F_{\alpha}(t)\Delta t + R_{\alpha}
\label{bd.eq}
\end{equation}

\noindent where $D^0$ is the short-time self-diffusion
coefficient, $\Delta t$ is the  time step,  and $R_{\alpha}$ is a
random displacement extracted from a Gaussian distribution with
zero mean and variance $2D^0\Delta t$. Taking  $\sigma$ as the
length unit and $\epsilon$ as the energy unit, $t_{BD}\equiv
\sigma^2/D^0$ becomes the natural time unit.

In both cases, the simulations were conducted with $N=1000$
particles in a cubic simulation box with periodic boundary conditions.
The initial configurations were generated using the
following procedure. First, particles were placed randomly in the
simulation box at the desired density, such that the maximum overlap
between particles was in the range $0.65\sigma-0.8\sigma$. To relax this
initial configuration and reduce or eliminate
the overlap between the particles we
tried two methods. In one of them we perform Monte Carlo cycles
\cite{swapmc} at a high temperature, and then decrease the temperature
for several steps until the original temperature was restored.
In the other method we uniformly expand the system by increasing
the length of the simulation box by a factor of at least $1.5$.
Then, we run MD or MC cycles while decreasing
the simulation box until the original value was reached.
We checked that the these two methods
produce equivalent results. Once the initial configuration
is constructed,
several thousand cycles are performed to lead the systems to
equilibrium, followed by at least two million cycles where
the data is collected.
In the case of MD simulations, temperature was kept constant
by simple rescaling of the velocities of the particles every 100 time
steps \cite{allen}.

Several structural and dynamic properties are calculated from
the equilibrium configurations generated in the simulations.
In particular, the radial distribution function $g(r)$ was
calculated using the standard approach \cite{allen}.
The static structure factor $S(k)$ can then be obtained as

\begin{equation}
S(k)=1 + 4\pi n\int [g(r)-1]\frac{\sin(kr)}{kr}r^2dr   .
\label{sdk.eq}
\end{equation}

\noindent
Alternatively, $S(k)$ can be calculated directly from the
positions of the particles in the simulation box \cite{hansen}.

Time correlation functions, like the mean squared displacement (MSD)

\begin{equation}
W(t)=\left< (\Delta \vec r(t))^2\right >/6  ,
\label{msd}
\end{equation}

\noindent
and the self-intermediate scattering function $F_S(k,t)$,

\begin{equation}
F_S(k,t) = \left < \frac{1}{N}\sum_{j=1}^{N}
\exp(-i\vec k \cdot \vec \Delta r_j(t))
\right >   ,
\label{fself.eq}
\end{equation}

\noindent
where $\Delta \vec r_j=\vec r_j(t)-\vec r_j(0)$,
were calculated using the efficient, low-memory algorithm
proposed in Ref. \cite{wdtblocks}.

Crystallinity of the systems was monitored through the
order parameters $Q_l$, especially $Q_6$,
defined as

\begin{equation}
Q_l=\left[ \frac{4\pi}{2l+1}\sum_{m=-l}^l |Q_{lm}|^2 \right]^{1/2}   ,
\end{equation}

\noindent
where  $Q_{lm}$ is basically the average, over all particles,
of the mean spherical harmonics $Y_{lm}(\hat r_{ij})$ established
between each particle $i$ and its close neighbors ($j=1,...,N_b(i)$),
where $N_b(i)$ is  the number of neighbors of the particle  \cite{qls}.
Since in this paper we are interested only in the amorphous
liquid state, when the simulations of monodisperse systems
exhibited crystalline order, thus indicating that the corresponding
volume fraction was beyond the freezing point, we discarded
that monodisperse run, and performed an alternative simulation
introducing size polydispersity to frustrate crystallization.
Polydispersity  is handled following previous
work \cite{gabriel}, where the diameters of the $N$ particles
are taken to be evenly distributed between
$\bar\sigma (1-w/2)$ and $\bar\sigma (1+w/2)$,
with $\bar\sigma$ being the mean diameter. We consider the case $w=0.3$,
corresponding to a polydispersity $P=w/\sqrt{12}=0.0866$.
Let us emphasize that this procedure was followed in both,
molecular and Brownian dynamics simulations, and that in
both cases only size polydispersity was introduced, leaving
all the other parameters unchanged (such as the mass or
the short-time self-diffusion coefficient of the particles).

At this point, it is important to emphasize that when the system
remains in its metastable liquid phase, the equilibration time
increases enormously as the system approaches its dynamic arrest
transition (see the detailed discussion in Ref. \cite{gabriel}).
This means that as the volume fraction increases in the metastable
region, the initial equilibration period will eventually be
insufficient, and will need to be adjusted to make sure that
the system indeed equilibrated properly, as recommended
in \cite{gabriel}. The present study, however, is not aimed
 at studying the equilibration process by itself, and hence,
we shall avoid approaching too closely the actual glass
transition, so as to focus our attention on the subject of this
work, namely, on the dynamic equivalence between soft-sphere
liquids.

\begin{figure}[htmb]
\centering
\includegraphics[width=7.0cm]{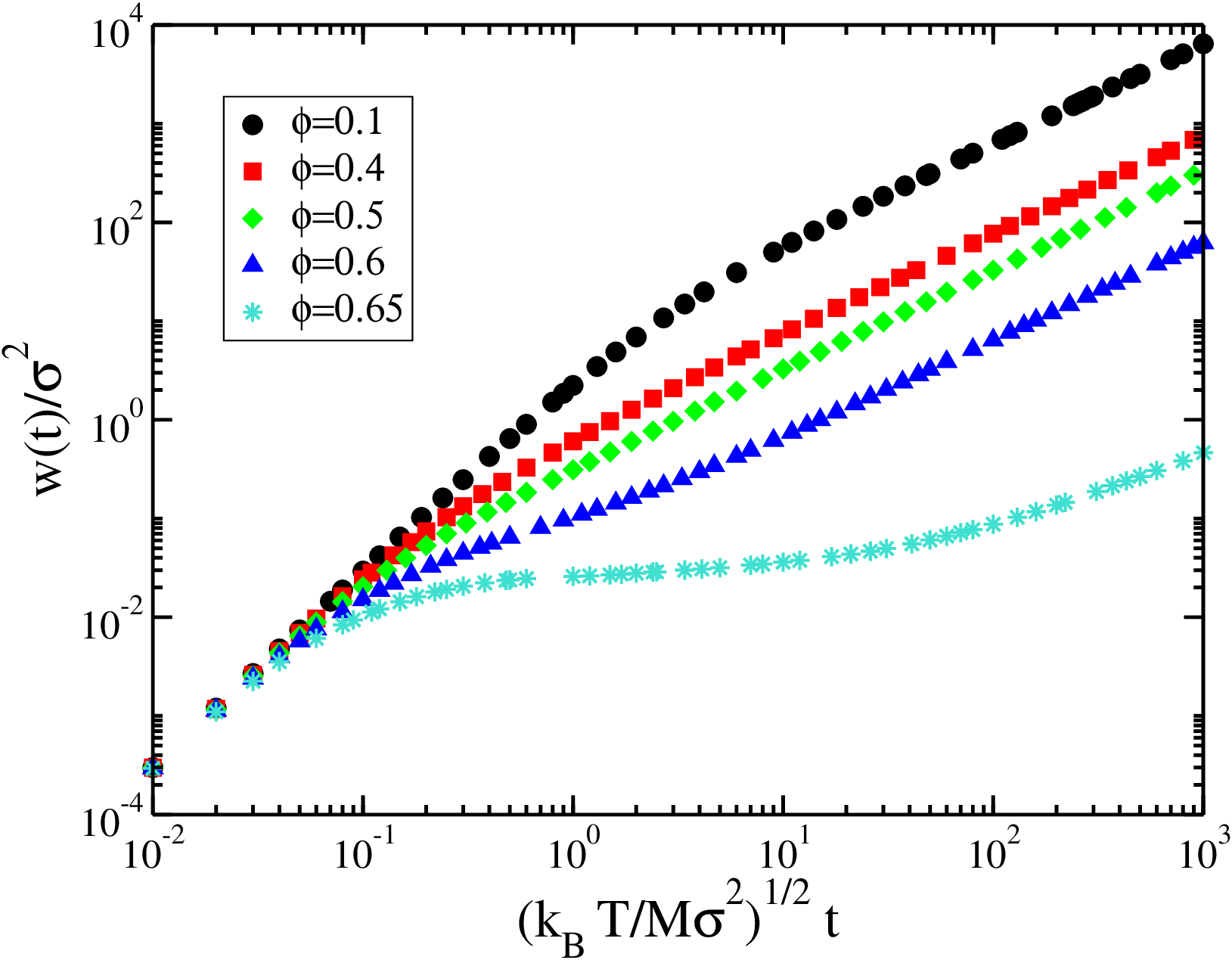}
\caption{
MSD from MD simulations TLJ systems with $\nu=15$, $T^*=1$, and the
volume fraction $\phi=0.1,0.4,0.5,0.6$, and 0.65.
The most concentrated sample ($\phi=0.65$)  is polydisperse.
}
\label{mdnu15_wdt.fig}
\end{figure}

\noindent
To illustrate the end result of this procedure,
in Fig. \ref{mdnu15_wdt.fig} we show the corresponding MSD for TLJ
system with $\nu=15$ at $T^*=1$ for volume
fractions $\phi=0.1,0.4,0.5,0.6,0.65$. These results fully cover the stable
fluid phase, with  the metastable liquid regime represented by the
polydisperse system with volume fraction $\phi\equiv
V^{-1}\sum_{i=1}^N \pi \sigma_i^3/6=0.65$ (whose monodisperse
counterpart is already highly ordered). In all samples the MSD
clearly exhibits the ballistic and diffusive time regimes typical
of atomic liquids.

\section{Soft-hard static equivalence}\label{sec3}

Although simulations are the main methodology employed here to
generate the static and the dynamic information of the model
systems above, the analysis of this information will rely on a few
theoretical notions, most notably the predicted static and dynamic
equivalence between soft-sphere and hard-sphere liquids. In this
analysis, however, we shall recurrently need the \emph{exact}
structural properties of the fluid of hard-spheres of diameter
$\sigma$ and volume fraction $\phi$, embodied in its RDF
$g_{HS}(r/\sigma;\phi)$ or in its static structure factor
$S_{HS}(k\sigma;\phi)$. For these structural properties a
virtually exact representation is provided by the Percus-Yevick (PY)
\cite{percusyevick,wertheim} approximation with its Verlet-Weis
correction, defined as \cite{verletweis}

\begin{equation}
g_{HS}(r/\sigma;\phi)= g^{(PY)}(r/\sigma_w;\phi_w),
\label{gdrpyvw}
\end{equation}

\noindent
and

\begin{equation}
S_{HS}(k\sigma;\phi)= S^{(PY)}(k\sigma_w;\phi_w),
\label{sdkpyvw}
\end{equation}

\noindent
with the parameters $\phi_w$ and $\sigma_w$ defined as

\begin{equation}
\phi_w\equiv \phi-\phi^2/16,
\label{phiw}
\end{equation}

\noindent
and

\begin{equation}
\sigma_w\equiv \sigma(\phi_w/\phi)^{1/3}.
\label{sigmaw}
\end{equation}

\noindent The functions $g^{(PY)}(x;\phi)$ and $S^{(PY)}(y;\phi)$
are  the solution of the Ornstein-Zernike equation with
PY closure for the HS fluid provided, for example, by
Wertheim \cite{wertheim} as easily programmable analytic
expressions. The resulting $g_{HS}(r/\sigma;\phi)$ will be
employed recurrently in the practical implementation of the
concept of \emph{static structural} equivalence between soft- and
hard-sphere systems. This notion was first introduced as an
essential aspect of the equilibrium perturbation theory of liquids
\cite{awc,verletweis,hansen}.

The equilibrium static structure of the  generic soft-sphere
system of the type discussed here is represented by the radial
distribution function (RDF) $g(r;n,T;\sigma,\epsilon,\nu)$, also
written in terms of dimensionless variables as
$g(r/\sigma;\phi,T^*,\nu)$, with $T^*\equiv k_BT/\epsilon$ and
$\phi = \pi n \sigma^3/6$. The physical notion behind the
principle of static equivalence is that at any state point
$(\phi,T^*,\nu)$, this soft-sphere system is structurally
identical to a hard-sphere system with a state-dependent effective
hard-sphere diameter $\sigma_{HS}$ and effective number density
$n_{HS}$. This means that for any state point $(\phi,T^*,\nu)$ one
can find a diameter $\sigma_{HS}=\sigma_{HS}(\phi,T^*,\nu)$ and a
number density $n_{HS}=n_{HS}(\phi,T^*,\nu)$ such that
$g(r;n,T;\sigma,\epsilon,\nu)\approx
g_{HS}(r;n_{HS},\sigma_{HS})$, where
$g_{HS}(r;n_{HS},\sigma_{HS})$ is the radial distribution function
of the HS system, also written as
$g_{HS}(r/\sigma_{HS};\phi_{HS})$, with $\phi_{HS} = \pi n_{HS}
\sigma_{HS}^3/6$. This condition for structural equivalence can
thus be written in terms of dimensionless variables as

\begin{equation}
g\left(\frac{r}{\sigma};\phi,T^*,\nu\right)
\approx g_{HS}(\frac{r}{\sigma_{HS}};\phi_{HS}),
\label{equivcond0}
\end{equation}

\noindent
with

\begin{equation}
\phi_{HS} =\frac{ \pi}{6} n_{HS} \sigma_{HS}^3= \lambda_n\lambda_\sigma^3\phi ,
\label{phihs}
\end{equation}

\noindent
where $\lambda_\sigma$ is just the state-dependent effective
hard-sphere diameter in units of $\sigma$,

\begin{equation}
\lambda_\sigma(\phi,T^*,\nu)\equiv \sigma_{HS}(\phi,T^*,\nu)/\sigma,
\label{lambdas}
\end{equation}

\noindent
and $\lambda_n$ is the state-dependent HS particle number
density in units of $n$,

\begin{equation}
\lambda_n(\phi,T^*,\nu)\equiv n_{HS}(\phi,T^*,\nu)/n.
\label{lambdan}
\end{equation}

\noindent Thus, the  condition for structural equivalence can be
written in scaled form as

\begin{equation}
g\left(\frac{r}{\sigma};\phi,T^*,\nu\right)\approx
g_{HS} \left(\lambda_\sigma^{-1}\frac{r}{\sigma};
\lambda_n\lambda_\sigma^3\phi\right).
\label{equivcond2}
\end{equation}

This equivalence condition can be used in several manners.  The
first one is to determine the parameters $\sigma_{HS}$ and
$n_{HS}$ that correspond to a given soft-sphere system at a given
state, i.e., to determine the functions
$\sigma_{HS}=\sigma_{HS}(\phi,T^*,\nu)$ and
$n_{HS}=n_{HS}(\phi,T^*,\nu)$. This might be done theoretically,
using specific assumptions. For example, one could assume that
$n_{HS}=n$ and that $\sigma_{HS}$ is $\phi$-independent, with
$\sigma_{HS}=\sigma_{HS}(T^*,\nu)$ determined by means of an
approximate version of the equivalence condition in Eq.
(\ref{equivcond0}). For example, the approximation employed in the
so-called blip-function method reads in general \cite{hansen}

\begin{equation}
\int _0^\infty 4\pi r^2 \left[ e^{-\beta u(r)} -
e^{-\beta u_{HS}(r)} \right]dr=0,
\label{blip0}
\end{equation}

\noindent
which for the TLJ system can be written as

\begin{equation}
\lambda^3_\sigma(T^*,\nu) = 1-3\int_0^1 dx x^2 { \exp
\left[-\frac{1}{T^*}(\frac{1}{x^{2\nu}}-\frac{2}{x^{\nu}}+1)
\right]}.
\label{blip1}
\end{equation}

\noindent
Evaluating $\lambda_\sigma (T^*,\nu)$ determines
$\sigma_{HS}$ as $\sigma_{HS}(T^*,\nu)=\sigma\lambda_\sigma (T^*,\nu)$.

Naturally, this or any other approximate scheme has a limited
range of validity. For example, as we shall see shortly, the
blip function method is reasonably accurate for finite-range,
moderately soft potentials, such as the TLJ liquid with softness
parameter $\nu \gtrsim 6$, but it fails completely for systems
with much softer and longer-ranged potentials, such as the HSY
fluid with  $K=554$, $z=0.149$, and $\phi$ of the order
of $10^{-4}$. Thus, it is important to search for a more robust
method to determine the functions
$\sigma_{HS}=\sigma_{HS}(\phi,T^*,\nu)$ and $n_{HS}=n_{HS}(\phi,T^*,\nu)$.

\begin{figure}
\begin{center}
\includegraphics[width=7.8cm]{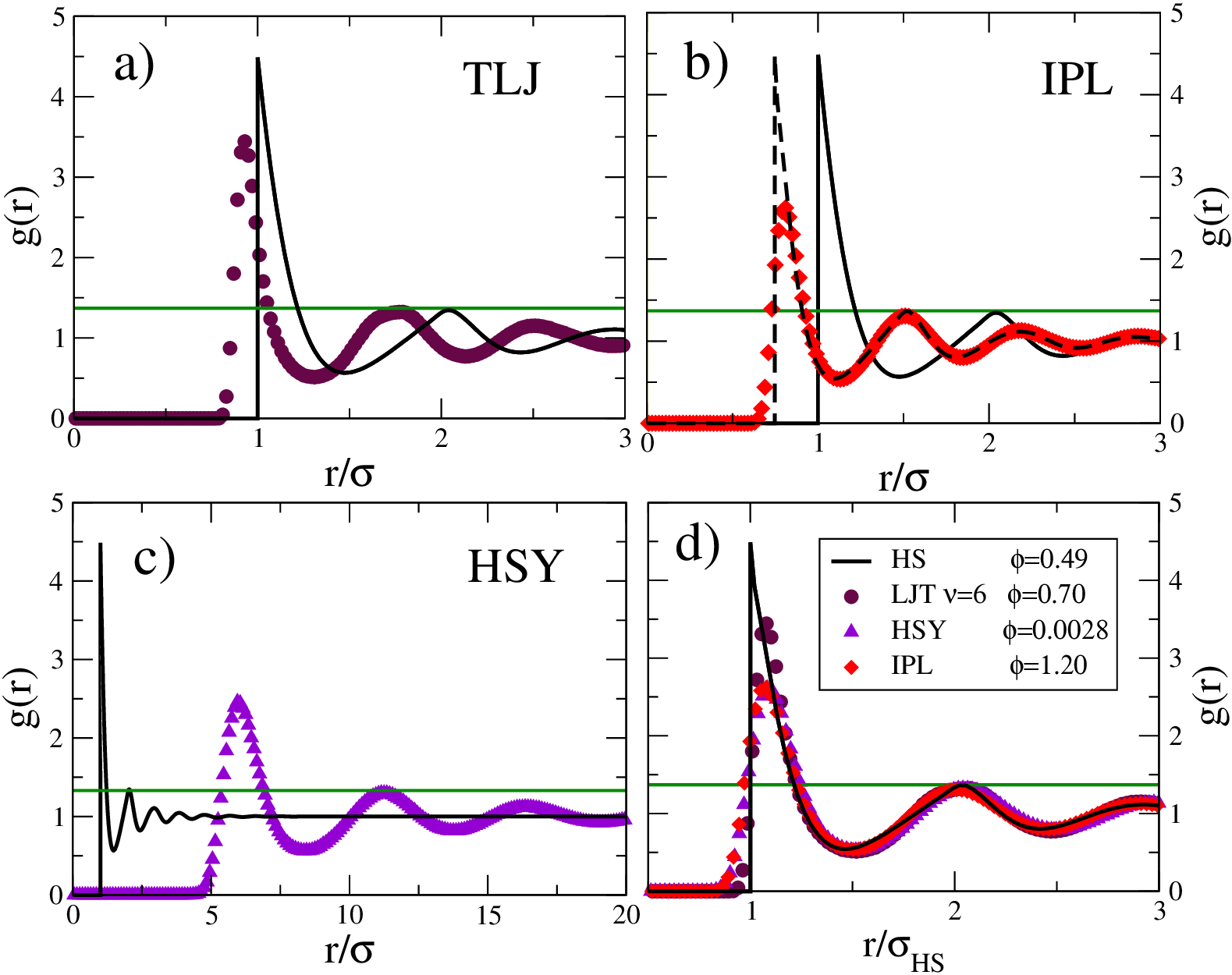}
\caption{ Radial distribution function for (a) a TLJ liquid with $\nu=6$, $T^*=1$, and $\phi=0.70$; (b) an IPL liquid with  $\nu=3$, $T^*=1$; and
$\phi=1.2$, and (c) a HSY liquid
with $K=554$, $z=0.149$, and $\phi=2.8 \times 10^{-3}$. All the systems share the same
$\phi_{HS}=0.49$. Solid lines corresponds to the
Percus-Yevick approximation for HS with the Verlet-Weis
correction. } \label{figescalamiento}
\end{center}
\end{figure}

One possible method, proposed in Ref. \cite{dynamicequivalence},
is to determine the diameter $\sigma_{HS}=\sigma_{HS}(\phi,T^*,\nu)$
and the number density $n_{HS}=n_{HS}(\phi,T^*,\nu)$
of the HS system whose radial
distribution function provides the best overall fit of the
exact RDF $g(r;n,T;\sigma,\epsilon,\nu)$ of the soft-sphere liquid
previously determined, for example, by computer simulations.
This method is illustrated here in Fig. \ref{figescalamiento},
where we plot simulation data for the RDF of three soft-sphere
model potentials (TLJ, IPL, and HSY). Fig.  \ref{figescalamiento}(a),
for example, plots the RDF $g(r/\sigma;\phi,T^*,\nu)$ of the
TLJ liquid with $\nu=6$, $T^*=1$, and $\phi=0.70$. Thus, we first
determine the effective hard sphere volume fraction $\phi_{HS}$
by plotting  the  exact RDF $g_{HS}(r/\sigma;\phi_{HS})$
of Eq. (\ref{gdrpyvw}) for various volume fractions until we
identify the value of $\phi_{HS}$ such that the height of its
second maximum matches the height of the second maximum of
the soft sphere RDF ($\approx$ 1.37, indicated by the thin
horizontal line in the figure). The solid curve is the
resulting hard sphere RDF.

This procedure assigns a unique value of $\phi_{HS}$ to that set
of values of the parameters  $(\phi,T^*,\nu)$, i.e., it determines
the function $\phi_{HS}=\phi_{HS}(\phi,T^*,\nu)$. As observed in
figure (\ref{figescalamiento}a), the height of these two second
maxima of $g(r)$ coincide, but their positions differ. One finds,
however, that a simple linear rescaling $r \to \lambda_\sigma^{-1}
r$ of the radial coordinate of this HS RDF, prescribed by the
equivalence condition in Eq. (\ref{equivcond2}), suffices to match
the position of both second maxima. This rescaling determines the
parameter $\lambda_\sigma$, and hence, also the effective
hard-sphere diameter $\sigma_{HS}$ at the state point
$(\phi,T^*,\nu)$ as $\sigma_{HS}=\sigma
\lambda_\sigma(\phi,T^*,\nu)$. Finally, the function
$n_{HS}=n_{HS}(\phi,T^*,\nu)$ is determined by

\begin{equation}
n_{HS}=n\left(\frac{\phi_{HS}}{\phi}\right)
\left(\frac{\sigma}{\sigma_{HS}}  \right)^3.
\label{nhs}
\end{equation}

\noindent
Following this procedure, in the illustrative example in
Fig.  \ref{figescalamiento}(a)  we find $\phi_{HS}=0.49$
and $\lambda_\sigma=\sigma_{HS}/\sigma= 0.88$; and therefore
 $\lambda_n =n_{HS}/n=1.01$. These numbers differ
only slightly from the results of the blip function method,
which assumes $\lambda_n =1$ and determines that
$\lambda_\sigma= 0.888$ and $\phi_{HS}=0.49$, a comparison that
illustrates the accuracy of the blip function method for the
TLJ potential with $\nu=6$. This accuracy improves for more
rigid potentials and deteriorates for softer and longer-ranged ones.

For example, Fig. \ref{figescalamiento}(b) reports an identical
exercise for the IPL potential with  $\nu=3$, $T^*=1$, and
$\phi=1.2$, whose RDF is represented by the symbols in the
figure. Here again the solid line is the RDF of the equivalent
HS system as a function of $r/\sigma$ and the dashed line
is the same RDF, but now plotted as a function of
$r/\sigma_{HS}=\lambda_\sigma^{-1} r/\sigma$,
to illustrate the overall agreement between the RDF of the
soft-sphere system and that of the equivalent HS system.
This method determines the effective HS parameters
$\phi_{HS}=0.49$, $\lambda_\sigma= 0.71$, and $\lambda_n =0.9679$.
In contrast, the  blip function method, which assumes
$\lambda_n =1$, determines  in this case the value
$\lambda_\sigma= 1.209$ and the unphysical HS volume
fraction $\phi_{HS}=2.41$.

Finally, Fig. \ref{figescalamiento}(c) reports the same exercise
but for a much softer and longer-ranged interaction, namely, the
HSY liquid with  $K=554$, $z=0.149$, and $\phi=2.8 \times
10^{-3}$. As before, the solid line is the RDF of the equivalent
HS system as a function of $r/\sigma$. In this case, the resulting
effective HS parameters are $\phi_{HS}=0.49$, $\lambda_{\sigma} =
5.55$, and $\lambda_n =1.0036$. In contrast, the  blip function
method ($\lambda_n =n_{HS}/n=1$) determines  the completely
unphysical values $\lambda_\sigma =\sigma_{HS}/\sigma= 27.6$ and
$\phi_{HS}=63.4$. Thus, the first conclusion of these three
illustrative examples is that the assumption that $\lambda_n
\approx 1$, employed in the blip function method above, may indeed
be a good approximation in the circumstances illustrated by these
three examples corresponding to the HS liquid at $\phi_{HS}=0.49$.
It is then the blip-function determination of the hard sphere
diameter through Eq. (\ref{blip1}) what is not an accurate
prescription.

Let us mention that in each of the three cases corresponding to
panels (a)-(c) of Fig.  \ref{figescalamiento} we chose to plot
the two equivalent RDFs as a function of the radial distance $r$
measured in the length unit $\sigma$ of the respective system.
This comparison, however, can also be done using instead the
effective HS diameter $\sigma_{HS}$ as the common unit length, as
it is done in Fig.  \ref{figescalamiento}(d). There we note, in
addition, that except for the shape of $g(r)$ near contact (which
is highly system-specific), the simulation data of the three
systems are actually coincident, and that we only have a single HS
RDF, corresponding to $\phi_{HS}=0.49$ and represented by the
solid line in Fig. \ref{figescalamiento}(d). This coincidence
illustrates another important feature, namely, that different
soft-sphere systems that share the same static structure also
share the same effective HS volume fraction.

This figure also illustrates the fact that the structural
equivalence condition in Eq. (\ref{equivcond0}) can also be used
in an inverse manner, i.e., to identify the state of a given
soft-sphere system whose structure matches the structure of a
prescribed HS system. In  reality,  what we actually did for each
of the three soft-sphere model systems in the examples in Fig.
\ref{figescalamiento} was to search for the state whose
structure matched the structure of the HS liquid with the
prescribed volume fraction $\phi_{HS}=0.49$. For this we varied
the soft-sphere density (or volume fraction $\phi$), keeping the
temperature fixed, until meeting this condition.

\begin{figure}
\begin{center}
\includegraphics[width=7.8cm]{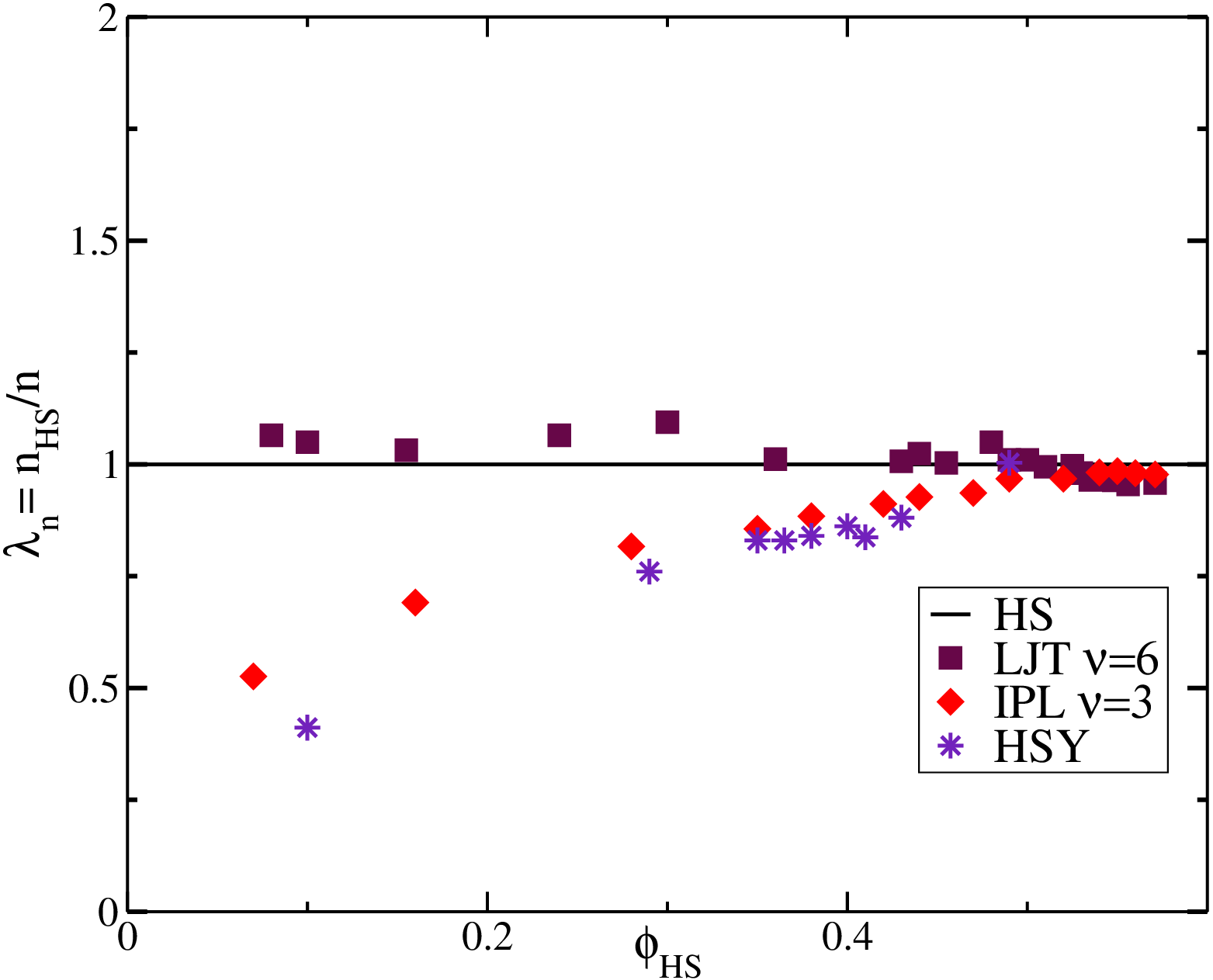}
\caption{
Ratio $n_{HS}/n$ as a function of the volume fraction of the equivalent
HS systems for LJT, IPL and HSY.
}
\label{nhsentren}
\end{center}
\end{figure}

In reality, the procedure just described to determine the
equivalent hard sphere system of a given soft-sphere model is not
limited to circumstances in which the condition $\lambda_n
=n_{HS}/n\approx 1$ is satisfied,  as in the last three examples.
For example, if we consider again the same systems discussed in
Fig. \ref{figescalamiento}, but at states that correspond to
effective hard-sphere volume fractions lower than $0.49$, the
structural equivalence will have the same  degree of accuracy as
the examples in Fig. \ref{figescalamiento}, even though the
condition $n_{HS}\approx n$ may definitely no longer be satisfied.
To illustrate the degree of the possible departures of the ratio
$n_{HS}/n$ from unity, in Fig. \ref{nhsentren} we plot
$n_{HS}/n$ for the TLJ, IPL, and HSY models, not as a function of
the respective volume fractions of each model system, but as a
function of the effective HS volume fraction, which is a common
indicator of the effective degree of packing of the three systems.
There we see that the TLJ system, whose pair interaction is always
short-ranged, virtually always satisfies the condition $\lambda_n
=n_{HS}/n\approx 1$, whereas the largest departures from this
condition are observed in the liquids with longer-ranged
potentials, such as the IPL and HSY systems at low effective
volume fractions.

\begin{figure}[htmb]
\centering
\includegraphics[width=7.0cm]{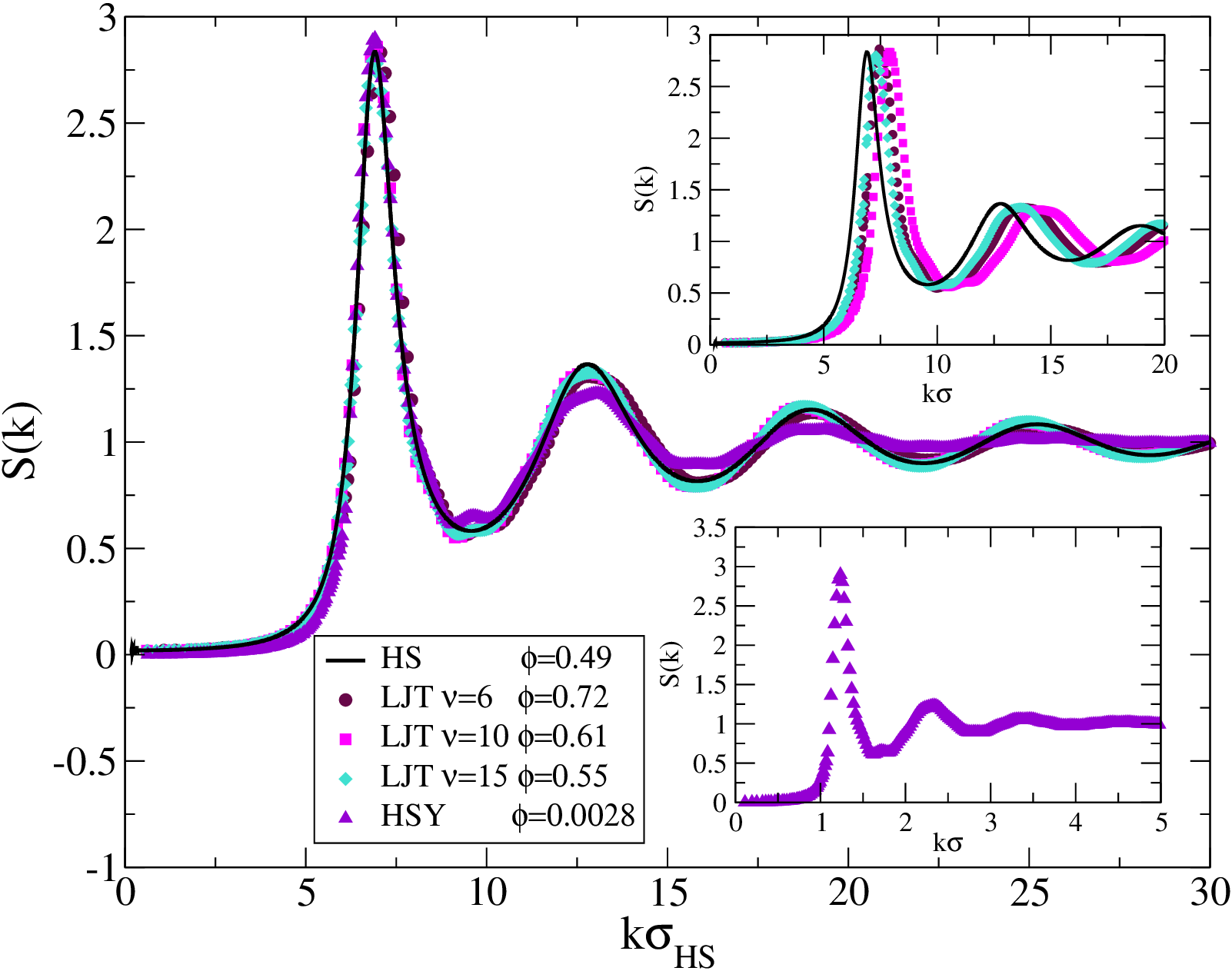}
\caption{Static structure factors for the three soft-sphere
systems of Fig. \ref{figescalamiento}.} \label{skmonos.fig}
\end{figure}

Another important observation is that for the model interaction
potentials employed in the present discussion, the height of
the second maximum of the RDF is not the only simple structural
order parameter. In reality, some other properties that derive
from the general equivalence condition in Eq. (\ref{equivcond0})
might serve as alternative  structural order parameters.
One of them is the main peak of the static structure factor
$S(k)$, whose height $S_{max}$ allows us to determine $\phi_{HS}$,
and whose rescaled position $k_{max}$ determines the effective
HS diameter $\sigma_{HS}$. This is illustrated in
Fig. \ref{skmonos.fig}, which exhibits the structure factors
of the same systems as in the previous figure, plotted as a
function, in one case of $k\sigma$ (insets) and in the other
case of $k\sigma_{HS}$ (main figure).

Since our study will include densities higher than the freezing
density of the monodisperse fluid, we need to introduce
polydispersity in our simulations and this requires us to adapt
the previously-described procedure to polydisperse systems.
Although polydispersity will not affect dramatically the most
relevant dynamic properties, it happens to have a profound effect
on the thermodynamic and structural properties, in  particular on
the height of the second peak of $g(r)$ and of the main peak of
$S(k)$. These structural order parameters are found to decrease
with polydispersity (for fixed volume fraction), and this requires
us to adapt the method described above, to identify in a simple
manner the effective hard-sphere system that corresponds to a
given \emph{polydisperse} soft-sphere liquid. The adapted
procedure is the following.  Consider a given soft-sphere system
with (size) polydispersity $P$ and mean diameter $\sigma$, whose
overall RDF $g(r/\sigma;\phi,T^*,\nu)$ has been measured or
simulated. Within a discretized representation the probability of
having a  diameter $\sigma_i$ ($i=1, 2,...,\nu$) is
$p(\sigma_i)=x_i$, with $x_i=n_i/n$ being the molar fraction of
species $i$. Under these circumstances $g(r/\sigma;\phi,T^*,\nu)$
is defined as $g(r/\sigma;\phi,T^*,\nu)\equiv
\sum_{\alpha,\beta=1}^{\nu} \sqrt{x_\alpha
x_\beta}g_{\alpha\beta}(r)$, with $g_{\alpha \beta}(r)$ being the
partial radial distribution functions. As in the monocomponent
case, determining the equivalent polydisperse hard-sphere system
whose overall RDF $g_{HS}(r/\sigma_{HS};\phi_{HS})$ matches the
simulated $g(r/\sigma;\phi,T^*,\nu)$, leads to the determination
of the total effective HS volume fraction $\phi_{HS}$ and the mean
HS diameter $\sigma_{HS}$.

To implement this procedure we need to determine the partial RDFs
of a multicomponent hard-sphere system at arbitrary total volume
fraction $\phi_{HS}$, but constrained to have the same
polydispersity $P$ as the soft-sphere system. For this, we
represent the equivalent polydisperse HS system  as an equi-molar
binary mixture of hard spheres of  diameters $\sigma_1 =
\sigma_{HS}(1-P)$ and $\sigma_2 = \sigma_{HS}(1+P)$ with
$\sigma_{HS}$ being the mean HS diameter. The overall RDF
$g_{HS}(r)$ of this system is given by $g_{HS}(r)\equiv
[g_{11}^{HS}(r)+2g_{12}^{HS}(r)+g_{22}^{HS}(r)]/2$, with
$g_{\alpha \beta}^{HS}(r)$ being the corresponding partial RDFs,
which  are obtained from the analytic solution of the
\emph{multicomponent} Percus-Yevick approximation
\cite{lebowitzhs2mixt,hiroike}, complemented again with the VW
correction \cite{williamsvanmegen}, i.e., $\phi_{HS}\to\phi_w$,
with $\phi_w\equiv \phi_{HS}-\phi_{HS}^2/16$, as in Eqs.
(\ref{gdrpyvw})-(\ref{sigmaw}), with $\sigma$ and $\phi$ reading
$\sigma_{HS}$ and $\phi_{HS}$. The resulting HS structure factors
will be denoted as PY-VW. As in the monocomponent case, the RDF
$g_{HS}(r;\phi_{HS})$ at arbitrary $\phi_{HS}$ is then compared
with the soft-sphere simulation results until determining the
value of $\phi_{HS}$ whose $g_{HS}(r;\phi_{HS})$ matches the
height of the second peak of the simulated RDF of the polydisperse
soft-sphere system.

\begin{figure}[htmb]
\begin{center}
\includegraphics[width=7.0cm]{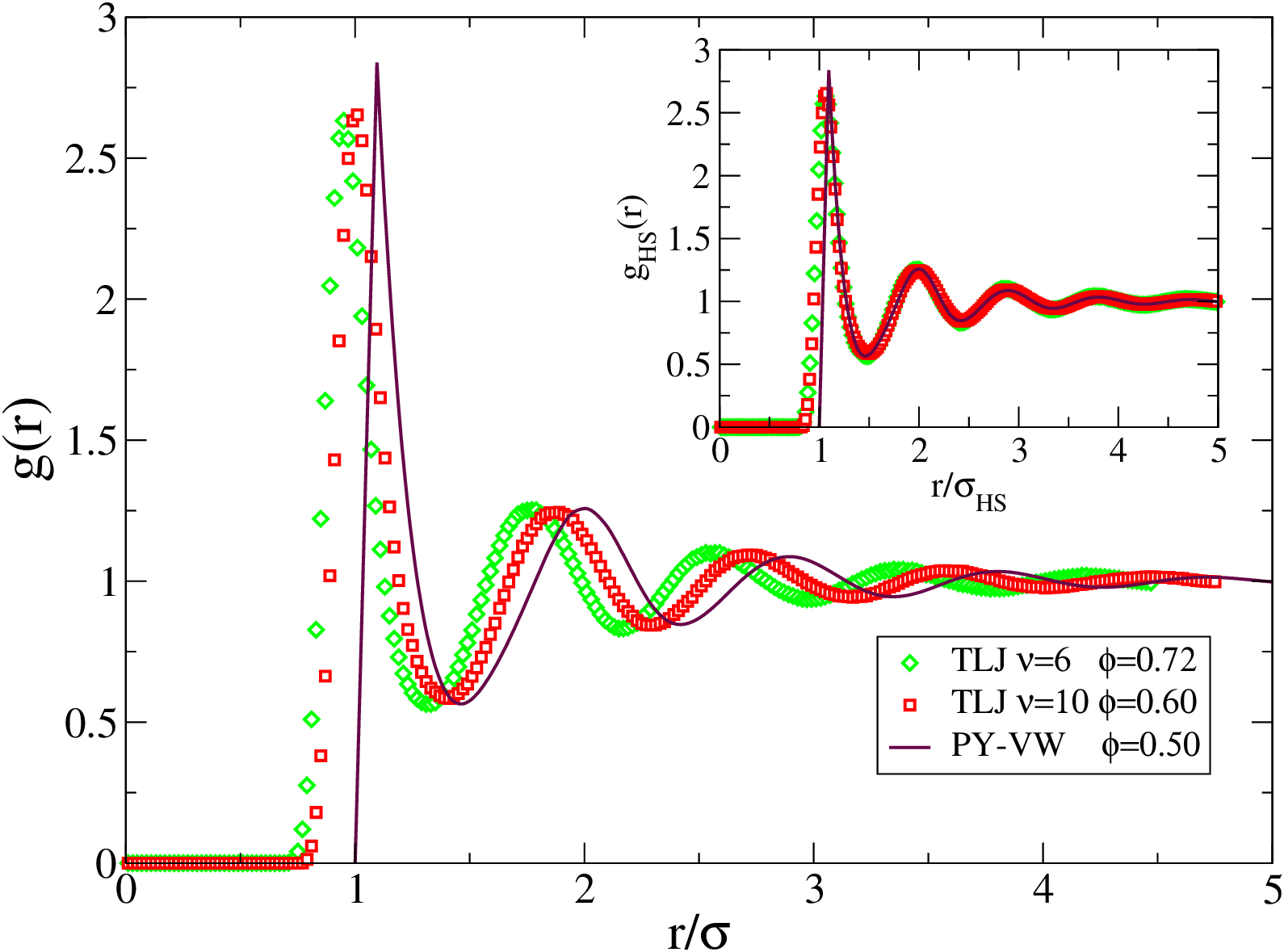}
\caption{
RDF's for polydisperse systems,  with $\nu=6$ and $\nu=15$ and  $\phi_{HS}=0.5$.
The line corresponds to the solution of the PY-VW theory for
the corresponding HS systems. Inset: RDFs
after scaling the radial distance.
}
\label{grpolis.fig}
\end{center}
\end{figure}

We illustrate this structural equivalence using a polydisperse
version of the TLJ model with different softness, $\nu=6$ and
$\nu=15$, but with densities corresponding to the same effective
HS volume fraction $\phi_{HS}=0.5$, and the same polydispersity
$P=0.0866$ (which is large enough to inhibit the crystallization
of hard spheres up to very high volume fractions \cite{gabriel}).
Fig. \ref{grpolis.fig} presents the simulation results for the
total RDF $g(r/\sigma;\phi,T,\nu)$, along with the theoretical
data for $g_{HS}(r;\phi_{HS})$ of the  corresponding HS binary
mixture. As  we can observe in the inset, the scenario is quite
similar to the one for monodisperse systems. Furthermore, this
inset also reveals that using the effective HS diameter to scale
the radial distance $r$ of the data in the main figure, collapses
the radial distribution function of the two polydisperse
soft-sphere systems  onto each other. Thus, the results in Figs.
\ref{figescalamiento} and \ref{grpolis.fig} show that our
protocol to determine the equivalent hard sphere system works very
well in a wide range of volume fractions for both monodisperse and
polydisperse systems. With this essential step covered, we now
investigate its implications on the dynamics of structurally
equivalent systems.

\section{Soft-hard Dynamic Equivalence}\label{sec4}
The {\it dynamic} extension of the previous soft--hard static
structural equivalence was discussed in
Refs. \cite{dynamicequivalence,soft1} in the context of the
dynamics of Brownian liquids, in which a short-time
self-diffusion coefficient $D^0$ describes the diffusive
microscopic dynamics of the colloidal particles
``between collisions". The following discussion also refers
to Brownian systems, but in the second part of this section
we shall consider atomic liquids, whose short-time dynamics is ballistic.

\subsection{Brownian liquids}\label{subsec4.a}

Although the dynamic equivalence we are about to discuss can
probably be understood from several perspectives, our original
insight and motivation derived from a straightforward prediction
of the self-consistent generalized Langevin equation (SCGLE)
theory of colloid dynamics. This theory can be summarized by a
closed system of equations for the collective and self
intermediate scattering functions $F(k,t)$ and $F_S(k,t)$
\cite{rmf,todos1,todos2}, which in Laplace space read

\begin{equation}
F(k,z) = \frac{S(k)}{z+\frac{k^{2}D^0S^{-1}(k)}{1+m(k) \Delta \zeta^*(z)}},
\label{fkz3}
\end{equation}

\noindent
and

\begin{equation}
F_S(k,z) = \frac{1}{z+\frac{k^{2}D^0}{1+m(k) \Delta \zeta^*(z)}},
\label{fskz3}
\end{equation}

\noindent
with $D^0$ being the short-time self-diffusion coefficient.
These equations become a closed system of equations when
complemented with the following approximate expression for
the time-dependent friction function $\Delta \zeta ^*(t)$,

\begin{equation}
\Delta \zeta^* (t) =\frac{D^0}{3\left( 2\pi \right) ^{3}n}\int d
{\bf k}\left[\frac{ k[S(k)-1]}{S(k)}\right] ^{2}F(k,t)F_{S}(k,t),
\label{dzdt}
\end{equation}

\noindent
and with the following definition of the ``interpolating"
function $m(k)$ \cite{todos2}

\begin{equation}
m(k) \equiv \frac{1}{1+\left(\frac{k}{k_c}\right)^\mu},
\label{lambdadk}
\end{equation}

\noindent with $\mu=2$ and with $k_c$ being the empirically chosen
cutoff wave-vector $k_c=1.118 k_{max}$, with $k_{max}$ being the
position of the main peak of $S(k)$.

From these equations and the condition for structural equivalence
in Eq. (\ref{equivcond0}),
$g\left(r/\sigma;\phi,T^*,\nu\right)\approx
g_{HS}(r/\sigma_{HS};\phi_{HS})$, it is not difficult to see that
when $\lambda_n =n_{HS}/n\approx 1$ (an excellent assumption in
many circumstances, such as those illustrated in Fig.
\ref{figescalamiento}), the dimensionless properties $F(k,t)$,
$F_S(k,t)$, and $\Delta \zeta ^*(z)$ of a given soft-sphere
system, can only depend on the wave-vector $k$ and the time $t$
through the dimensionless variables $k\sigma_{HS}$ and
$D^0t/\sigma_{HS}^2$. Furthermore, scaled in this manner, Eqs.
(\ref{fkz3})-(\ref{lambdadk}) above become identical to those of
the hard-sphere system at volume fraction $\phi_{HS}$. This
implies the existence of the dynamic equivalence summarized by the
statement that the dynamic properties, such as the self
intermediate scattering function (self-ISF)
$F_S(k,t;n,T;\sigma,\epsilon,\nu; D^0)$, of the fluid with soft
repulsive potential $u(r)$, can be approximated by the
corresponding property of the (statically) equivalent hard-sphere
Brownian liquid whose particles diffuse with the same $D^0$, i.e.,
$F_S(k,t;n,T;\sigma,\epsilon,\nu; D^0) \approx
F^{(HS)}_S(k,t;n,\sigma_{HS}; D^0)$. This relationship can be
written in terms of dimensionless variables as
\begin{equation}
F_S(k\sigma,D^0t/\sigma^2;\phi,T^*,\nu)
\approx F^{(HS)}_S(k\sigma_{HS},D^0t/\sigma_{HS}^2;\phi_{HS}).
\label{scalingforfself}
\end{equation}
Some consequences of this dynamic equivalence were illustrated in
Refs. \cite{soft1} and \cite{dynamicequivalence}
in the context of the TLJ potential. Those references, however,
discussed in detail only the limit of moderate softness
($\nu \gg 1$), in which the strong similarity with the HS
potential leads to the additional simplification that
$\sigma_{HS}(n,T,\nu)$ becomes $n$-independent, and given
by the ``blip function" approximation \cite{hansen,soft1}.
These, however, are actually unessential restrictions,
as illustrated  by the Brownian dynamics
simulations for the IPL and HSY models discussed below.

The universality summarized by Eq. (\ref{scalingforfself})
leads to the corresponding scaling rules for other properties.
For example, let
\begin{equation}
W(t;T^*,\phi,\nu)\
\equiv <(\Delta \textbf{r}(t))^2>/6
\label{msdaswdt0}
\end{equation}
be the mean squared displacement (MSD) of any soft-sphere
liquid at a given state $(T^*,\phi,\nu)$ that structurally
maps onto the hard-sphere liquid of diameter
$\sigma_{HS}(T^*,\phi,\nu)$ and volume fraction
$\phi_{HS}(T^*,\phi,\nu)$. Then the normalized MSD
\begin{equation}
W^*(t^*;T^*,\phi,\nu)\
\equiv <(\Delta \textbf{r}(t^*))^2>/6\sigma^2_{HS}(T^*,\phi,\nu),
\label{msdaswdt}
\end{equation}
with $t^*\equiv D^0t/\sigma_{HS}^2$, will be identical
to that of the equivalent hard-sphere fluid,
\begin{equation}
W^*(t^*;T^*,\phi,\nu)=
W^*_{HS}[t^*; \phi_{HS}(T^*,\phi,\nu)],
\label{scaledmsd}
\end{equation}
and for that matter, to that of any other soft-sphere
liquid that is structurally equivalent to the HS system
with the same volume fraction $\phi_{HS}$.

\begin{figure}[htmb]
\begin{center}
\includegraphics[width=7.0cm]{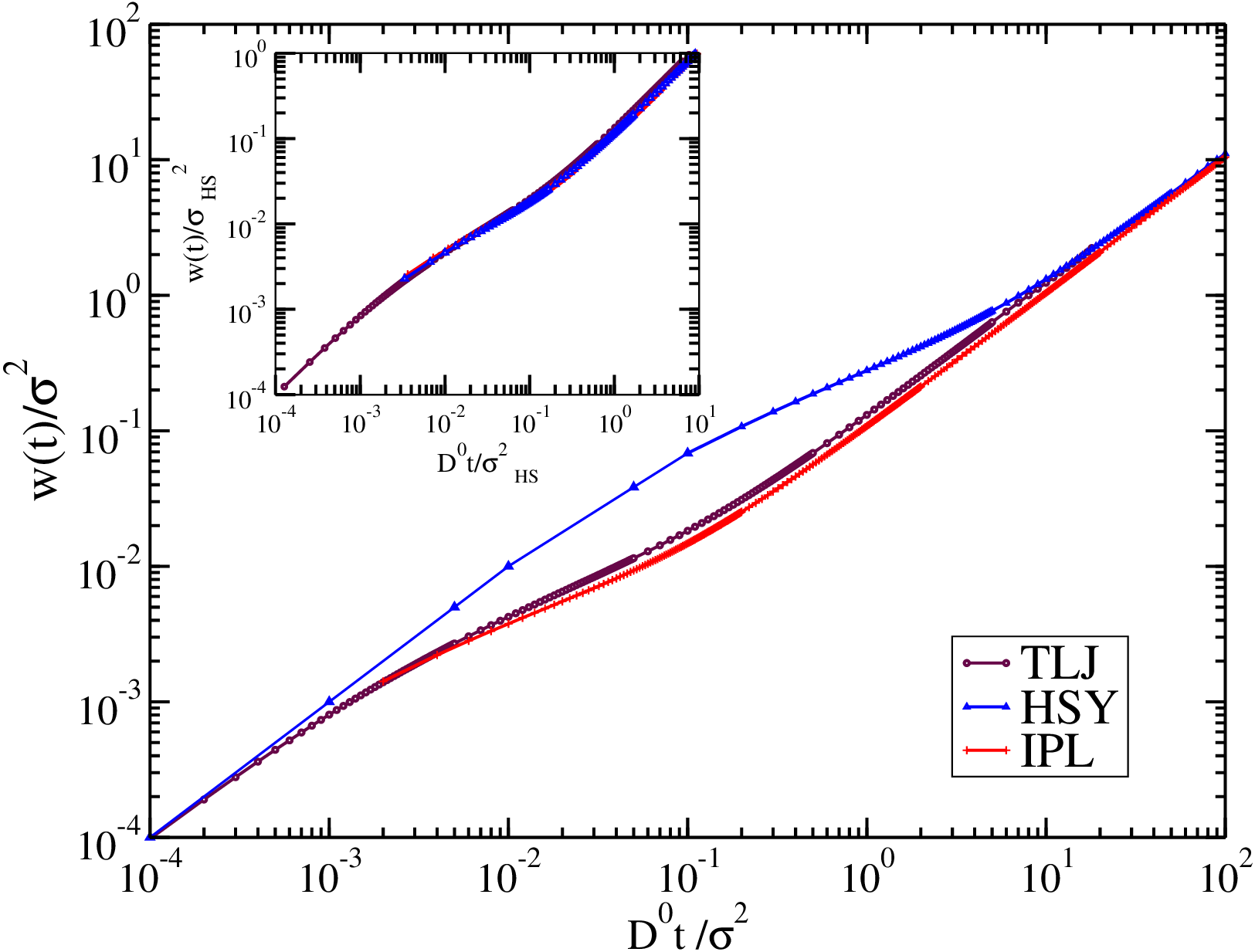}
\caption{ BD results for the MSD of the systems
in Fig. \ref{figescalamiento}, in original units (main panel), and scaled with the hard-sphere
unit. } \label{wdtbd.fig}
\end{center}
\end{figure}

To test this prediction in Fig. \ref{wdtbd.fig}  we present the
BD results for the mean squared displacement of the three
soft-sphere systems discussed in Fig. \ref{figescalamiento},
i.e., the TLJ with $\nu= 6$, the IPL with $\nu= 3$ (both at
$T^*=1$), and the HSY with $K=554$ and $z=0.149$, all of them
corresponding to an equivalent volume fraction $\phi_{HS}= 0.49$.
The MSD is presented in the figure in the natural units of the BD
simulations, i.e. as $[W(t)/\sigma^2]$ vs. $[D^0t/\sigma^2]$. We
observe that the MSD exhibits the two linear regimes typical of
Brownian systems \cite{pusey0}: at short times $[W(t)/\sigma^2]
\approx [D^0t/\sigma^2]$ whereas at long times $[W(t)/\sigma^2]
\approx  [D_Lt/\sigma^2]$, where  $D_L$ is the \emph{long-time}
self-diffusion coefficient. Thus, at short times the MSD must be
the same for all systems and states, a condition clearly fulfilled
by the data plotted in the figure.

In the long-time regime, on the other hand, the MSD reads
$[W(t)/\sigma^2] \approx D^* [D^0t/\sigma^2]$, i.e., it is
proportional to the scaled long-time self-diffusion coefficient
$D^*$, defined as
\begin{equation}
D^*\equiv D_L/D^0.
\label{dstar}
\end{equation}
This property does depend on the interparticle
interactions, and hence, on the particular  system and on
its state, i.e.,  $D^*=D^*(T^*,\phi,\nu)$. Thus, the various $W(t)$ in
Fig.  \ref{wdtbd.fig} should differ in their long-time
behavior. They, however, exhibit the same long-time limit.
The reason for this is that the
dynamic equivalence condition in  Eq. (\ref{scalingforfself})
implies that the dimensionless parameter $D^*$ depends on
$(T^*,\phi,\nu)$ only through the effective HS volume
fraction $\phi_{HS}=\phi_{HS}(T^*,\phi,\nu)$,
\begin{equation}
D^{*}(T^*,\phi,\nu) \approx
D^{*}_{HS}[\phi_{HS}(T^*,\phi,\nu)],
\label{scalingfordsl}
\end{equation}
and  the three systems in Fig.  \ref{figescalamiento} were
chosen to have the same  $\phi_{HS}\ (=0.49)$. Thus, they share
the same value of $D^* (\approx 0.1)$.

In fact, for the very same reason (see the scaling in Eq.
(\ref{scaledmsd})) these three systems must actually share the
full dependence of $W^*(t^*)$ on the scaled time $t^*\equiv
D^0t/\sigma_{HS}^2$, i.e., the three different MSDs in Fig.
\ref{wdtbd.fig} should collapse onto the same curve when plotted
as a function of $t^*$. This is indeed what we find, as
illustrated in the inset of this figure. Furthermore, according to
Eq. (\ref{scaledmsd}), the resulting master curve then determines
the function $W^*_{HS}[t^*; \phi=0.49]$. The SCGLE theory  for
Brownian systems (i.e., Eqs. (\ref{fkz3})-(\ref{lambdadk})),
besides predicting this scaling also provides an approximate
prediction for this function. The results of the SCGLE theory
applied to the HS fluid  follow closely the simulations results.

\begin{figure}[htmb]
\centering
\includegraphics[width=6.0cm]{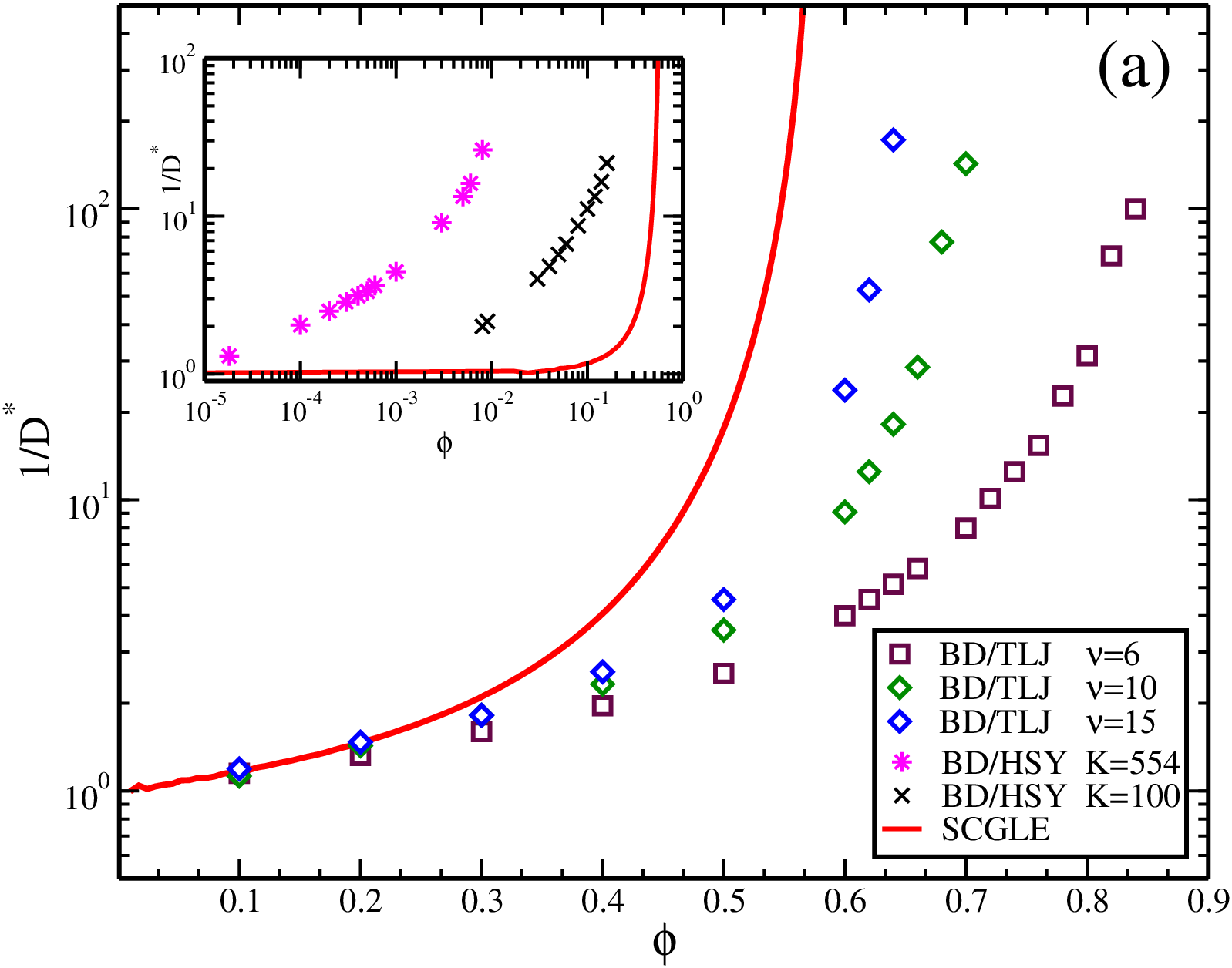}
\includegraphics[width=6.0cm]{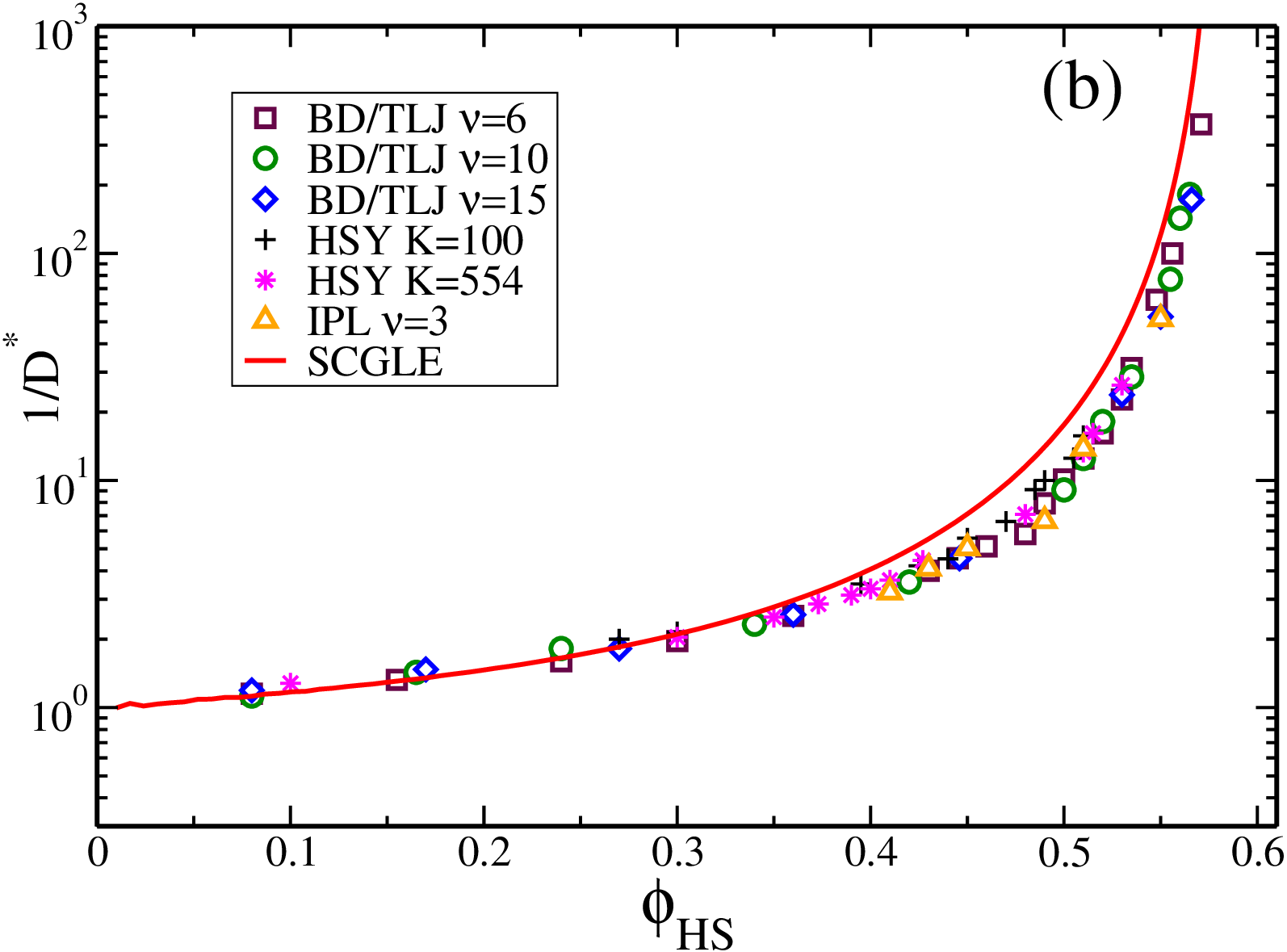}
\caption{
Inset (a): $D^*=D_L/D^0$ from
DB simulations for TLJ systems with $\nu=6$, $\nu=10$ and $\nu=15$,
at different volume fractions $\phi$.
Inset: Results for two HSY systems
($K=554$, $z=0.149$, and $K=100$ $z=1.0$). Inset (b): Results plotted
as a function of $\phi_{HS}$.Data for the IPL($\nu=3$) is included.
The solid line represents the solution for HS
systems from the colloidal SCGLE theory.
}
\label{diffbd.fig}
\end{figure}

The results presented in this  figure are concerned with the full
MSD of three model systems that share the same effective volume
fraction $\phi_{HS}=0.49$. Let us next extend our study  to a
wider range of effective HS volume fractions, focusing on
long-time properties such as the long-time self-diffusion
coefficient $D_L$ and the $\alpha$-relaxation time
$\tau_{\alpha}$. We start by presenting in Fig.
\ref{diffbd.fig}(a) the BD results for the inverse of
$D^*=D_L/D^0$ as a function of the respective volume fraction
$\phi$ for several soft-sphere  systems. The main panel of the
figure shows how the simulated $D^*$ of three TLJ systems depends
on volume fraction and on softness. For  instance, $D^*$ is, as
expected, a decreasing function of $\phi$ and $\nu$, and in the
low-$\phi$ limit all the results converge to the correct limiting
value $D^*(\phi\to 0)=1$. For reference, in this figure we include
the approximate SCGLE prediction of the function $D^*_{HS}(\phi)$
corresponding to the HS fluid (the solid line in the figure). It
is clear that as the potential becomes stiffer the function
$D^*(\phi)$ gradually approaches the function $D^*_{HS}(\phi)$.

Model systems with longer-ranged interactions can  depart even
further from this HS limit. Thus, in the inset of the same figure
we compare the results for $D^{*}(\phi)$ of two HSY systems with
the predicted HS limit (also represented by the solid line). This
comparison exhibits a much more pronounced difference  compared
with the  shorter-ranged systems in the main panel. Here the
values of $D^*$ in the liquid regime of the HSY systems correspond
to volume fractions in the range $10^{-5}-10^{-1}$ (from weakly to
highly structured conditions), whereas the relevant volume
fractions of the TLJ systems fall in the typical range
$0.1\lesssim \phi\lesssim 1$. Despite these differences, however,
when these data, as well as the data for the TLJ systems in the
main figure, are plotted as suggested by the scaling in Eq.
(\ref{scalingfordsl}),  all of them fall on a well-defined master
curve, as demonstrated in Fig. \ref{diffbd.fig}(b). This master
curve must then determine the exact hard-sphere function
$D^{*}_{HS}[\phi]$. The solid line in the figure is the
corresponding approximate SCGLE prediction for this function,
whose comparison with the exact master curve defined by the
collapsed simulation data indicates the level of quantitative
accuracy of this theory.

Let us notice that, although the results in the figure  illustrate
the dynamic equivalence at the very particular condition
$t\rightarrow \infty$, our simulations show that basically the
same scaling holds at all times for time-dependent properties such
as the scaled time-dependent self-diffusion coefficient
$D^*(t)\equiv W(t)/6D^0t$ (illustrated in Fig. 4 of Ref.
\cite{dynamicequivalence})), or the scaled MSD $W^*(t^*)$,
illustrated here in Fig.  \ref{wdtbd.fig} with the results of
three model systems that meet the condition $\lambda_n
\approx 1$ at an effective volume fraction is
$\phi_{HS}=0.49$. We have extended this study, however, to the IPL
and HSY potentials at low effective HS volume fractions, the
regime in which appreciable deviations from the condition
$\lambda_n $ are exhibited (see Fig. \ref{nhsentren}).
The corresponding scaled results for $D^*(\phi_{HS})$, presented
in Fig. \ref{diffbd.fig}(b), demonstrate that the predicted
dynamic equivalence has a wide range of validity, requiring only
the static structural equivalence discussed in Sec. \ref{sec3},
but not necessarily the condition $n_{HS}\approx n$, in spite of
the fact that our original insight of the soft-hard dynamic
equivalence derived from the structure of the SCGLE equations
within the condition $\lambda_n =n_{HS}/n\approx 1$.

\begin{figure}[htmb]
\centering
\includegraphics[width=7.0cm]{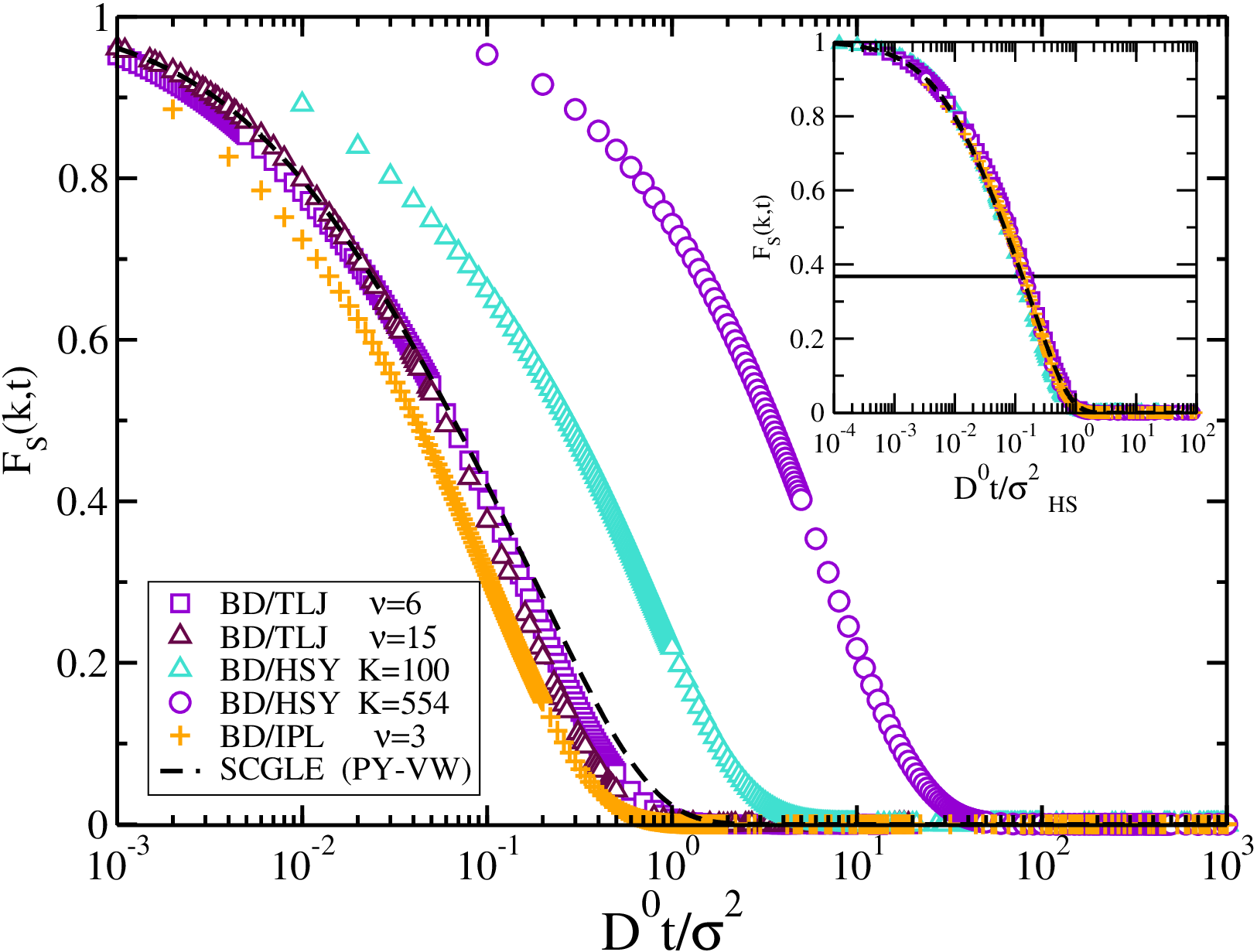}
\caption{
$F_S(k_{max},t)$ from BD
simulations for two TLJ systems with $\nu=6$ and $\nu=15$, IPL system,
and two Yukawa systems . All systems
have an effective volume fraction $\phi_{HS}=0.49$.
In the inset the results are presented as a function of $D^{0}t/\sigma_{HS}^{2}$ and $k\sigma_{HS}$, with $k=k_{max}$.
}
\label{fs_bd.fig}
\end{figure}

Let us now turn our attention to the relaxation of the
intermediate scattering function $F_S(k,t)$. To exhibit the
dynamic equivalence predicted by Eq. (\ref{scalingforfself}), we
evaluate $F_S(k,t)$ for two TLJ and two HSY systems at states that
share the same equivalent volume fraction, $\phi_{HS}=0.49$. In
the main frame of Fig. \ref{fs_bd.fig} we plot  $F_S(k_{max},t)$
as a function of the scaled time $D^0t/\sigma^2$, a format in
which the results for the TLJ and HSY systems differ notoriously.
In particular, the $\alpha$-relaxation time $\tau_{\alpha}$ of
these systems, defined by the condition
$F_S(k,\tau_{\alpha})=1/e$, differ by more than one decade.
According to Eq. (\ref{scalingforfself}), however, the same
results should collapse onto a single master curve upon the
transformation to HS units, and provided that $F_S(k,t)$ for the
various systems is evaluated at structurally equivalent
wave-vectors (i.e., same value of $k\sigma_{HS}$). To meet this
iso-structural requirement for the three systems in the figure we
have evaluated $F_S(k,t)$ at the position $k_{max}$ of the
corresponding static structure factors. The resulting master curve
is shown in the inset of the figure, in which the data are plotted
as a function of the scaled time $D^0t/\sigma_{HS}^2$. Such
scaling of $F_S(k,t)$, in its turn, leads to identical
$\alpha$-relaxation times for iso-structural systems, when
expressed in the new units, regardless of the softness or range of
the interaction between the particles.

\begin{figure}[htmb]
\centering
\includegraphics[width=7.0cm]{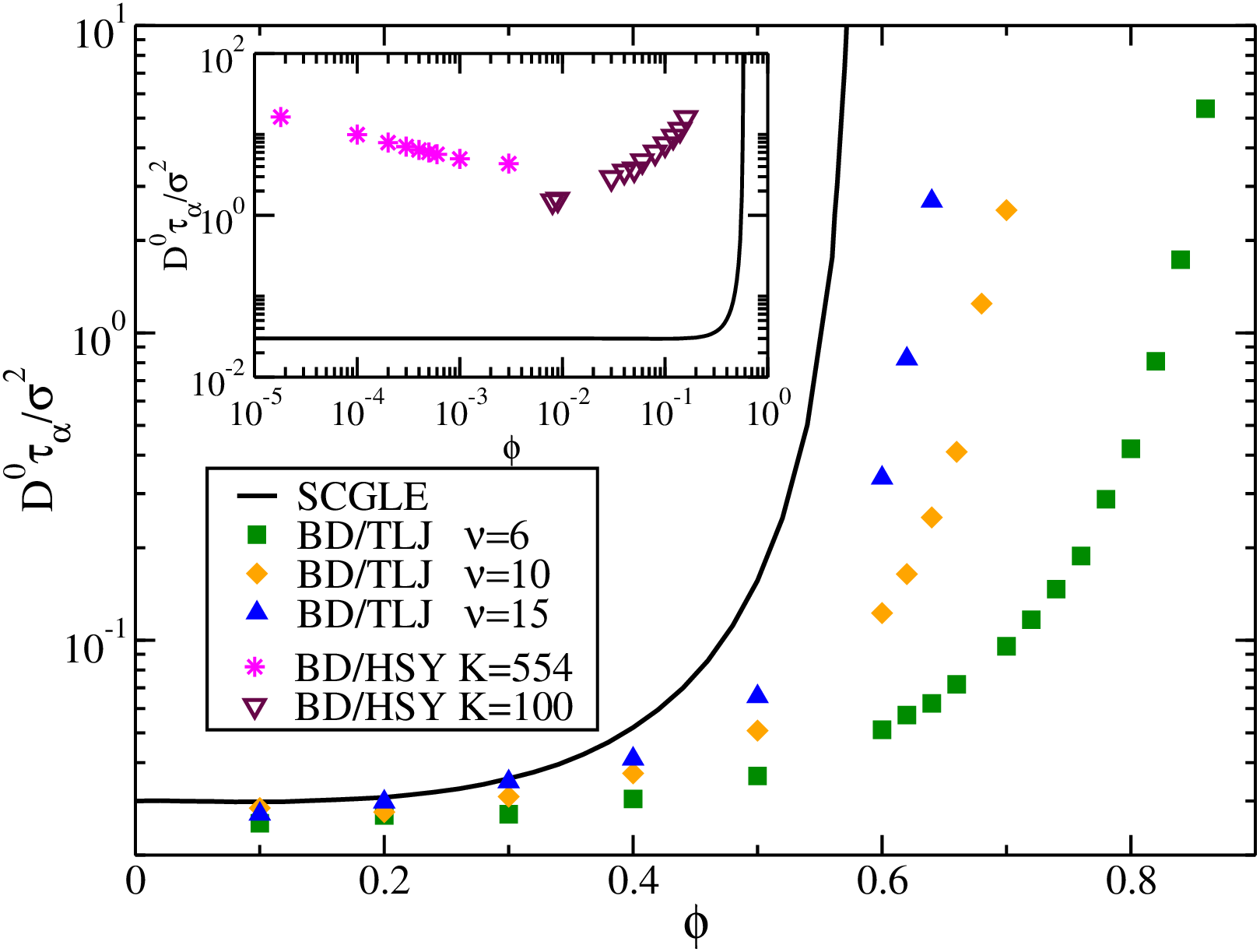}
\caption{
$\tau_{\alpha}(\phi)$, in BD units,
for  TLJ systems. The solid line represents the
SCGLE results for HS systems.
Inset: Results for two HSY systems.
}
\label{tau.fig}
\end{figure}

Another manner to exhibit this dynamic equivalence  starts with
the  BD simulations of the $\alpha$-relaxation time, for
several soft-sphere systems in an extended range of volume
fractions, evaluated at a structurally identical wave-vector (we take $k=k_{max}$).
In the main panel of Fig. \ref{tau.fig} we present
the unscaled results for $\tau_{\alpha}$ corresponding to the TLJ
systems with  $\nu=6$, $\nu=10$ and $\nu=15$. There we observe
that $\tau_{\alpha}$ exhibits the typical monotonic slowing down
with concentration, increasing faster at higher $\phi$, at a rate
that depends strongly on the softness of the potential, following
a similar pattern as $1/D^*$ in the mainframe of Fig.
\ref{diffbd.fig}(a). Here too, increasing $\nu$  leads to results
progressively closer to the curve predicted by the SCGLE theory
for the HS system, represented in the figure by the solid line.

In contrast, the corresponding simulation data  of $\tau_{\alpha}$
for the longer-ranged HSY  systems, presented in the inset of Fig.
\ref{tau.fig}, exhibit a notoriously different
$\phi$-dependence. The difference is mainly quantitative in the
case of the HSY system with $K=100$ and $z=1.0$, since the results
for $\tau_{\alpha}$ presented in the figure also increase
monotonically, although at much smaller values of $\phi$ with
respect to the hard-sphere system (represented again by the solid
line). In the case of the system  with $K=554$ and $z=0.149$,
however, the corresponding differences seem to be even
qualitative, since the data presented change non-monotonically
with  $\phi$. Although this contrast may appear dramatic, however,
it actually reflects a rather trivial consequence of the facts
that at low volume fractions $\tau_{\alpha}\approx1/k^2D^0$ and
that for this long-ranged, strongly-interacting, HSY system,
$k_{max}\approx 2\pi/n^{-1/3}$; thus, at low volume fractions
$\tau_{\alpha}\propto \phi^{-2/3}$. Thus, this qualitative feature
would be absent if we had plotted  $\tau^*\equiv
k^2D^0\tau_{\alpha}$, rather than the unscaled $\alpha$-relaxation
time.

\begin{figure}[htmb]
\centering
\includegraphics[width=7.0cm]{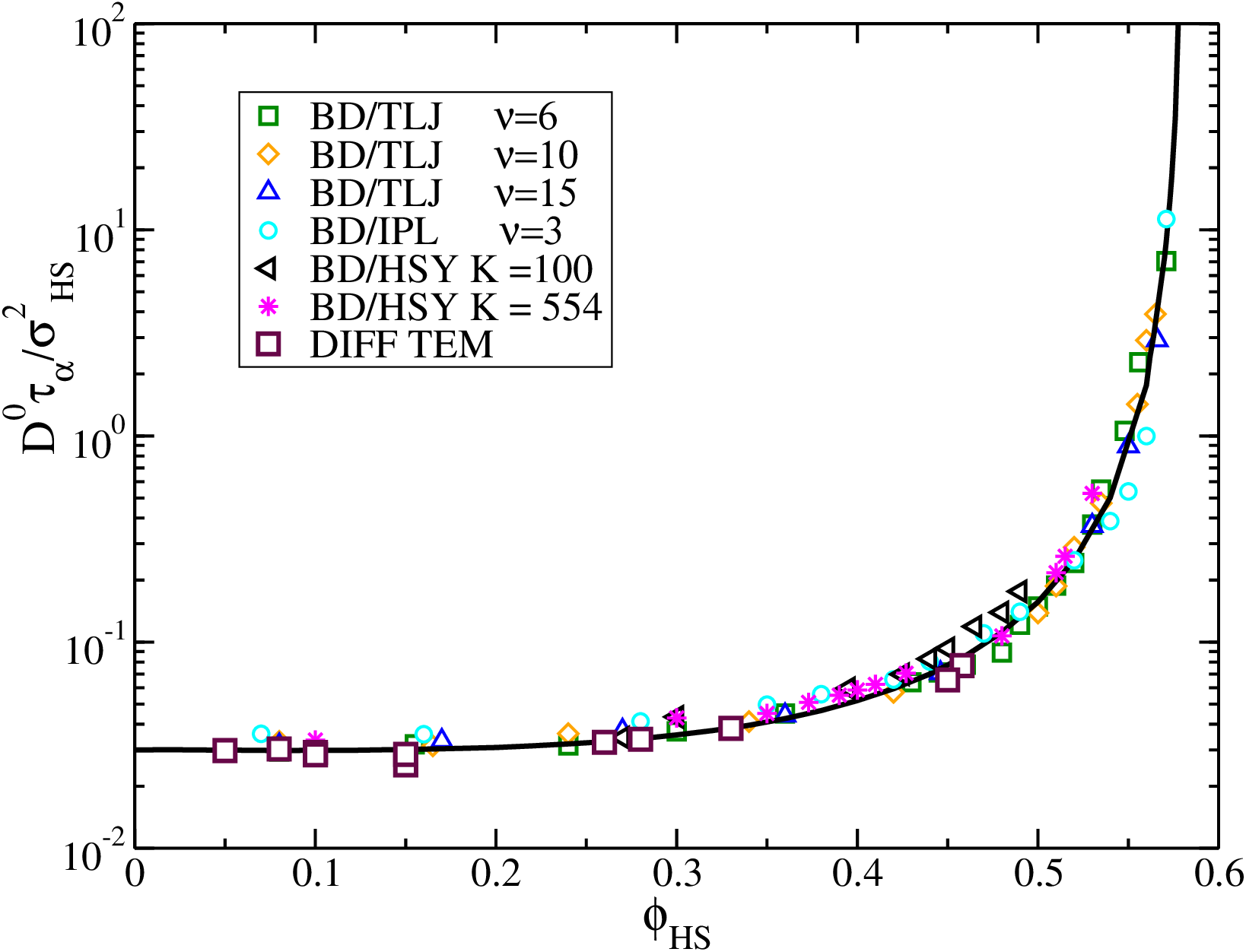}
\caption{
$\alpha$-relaxation time $\tau_{\alpha}$, in HS units, as
a function of $\phi_{HS}$.
The solid line corresponds to the results from the SCGLE theory
applied to HS fluids.
}
\label{tau_bd_hs.fig}
\end{figure}

This scaling, however, will only remove the apparently  anomalous
$\phi$-dependence of $\tau_{\alpha}$, but not the quantitative
difference in the range of volume fractions at which the sharp
increase of $\tau_{\alpha}$ occurs. However, the dynamic  scaling
in Eq. (\ref{scalingforfself}), illustrated in Fig.
\ref{fs_bd.fig} with $F_S(k,t)$ for the case $\phi_{HS}=0.49$,
should now align the data for $\tau_{\alpha}(\phi)$ in Fig.
\ref{tau.fig} with those of the HS system upon the
transformation to HS units. For this we mean to plot
$\tau_{\alpha}$ scaled as $D^0 \tau_{\alpha}/\sigma_{HS}^2$, as a
function of $\phi_{HS}(\phi,T^*)$. This is done in Fig.
\ref{tau_bd_hs.fig}, where we  corroborate that indeed the
transformed data fall in a master curve that follows closely the
solid line, i.e., the SCGLE theoretical predictions for the HS
fluid.

\subsection{Atomic liquids}\label{subsec4.b}

Let us now discuss this dynamic equivalence in the context  of
atomic liquids. For this, let us notice that in order to discus
the dynamic equivalence between soft- and hard-sphere Brownian
fluids we first unambiguously defined an effective hard-sphere
volume fraction and diameter, $\phi_{HS}=\phi_{HS}(\phi,T^*)$  and
$\sigma_{HS}=\sigma_{HS}(\phi,T^*)$, for each state $(\phi,T^*)$
of the soft-sphere system. Then, the dynamic equivalence was
simply exhibited by expressing  the dimensionless properties of
the system not in terms of the natural units of the soft system,
namely, $\phi$, $\sigma$, and $t^0\equiv \sigma^2/D^0$, but in
terms of the units of the equivalent HS system, namely,
$\phi_{HS}$, $\sigma_{HS}$, and $t^0_{HS}\equiv
\sigma^2_{HS}/D^0=\lambda^{2}_{\sigma}t^0$. Since the definition
of the effective hard-sphere diameter only involves the comparison
of the equilibrium static structure of the soft-sphere system with
the corresponding structure of the equivalent hard-sphere system
(see Eq. (\ref{equivcond0})), it is natural to expect that the
dynamic equivalence discussed above in the context of Brownian
systems also holds independently of the underlying microscopic
dynamics, i.e., also for atomic liquids.

\begin{figure}[htmb]
\begin{center}
\includegraphics[width=7.0cm]{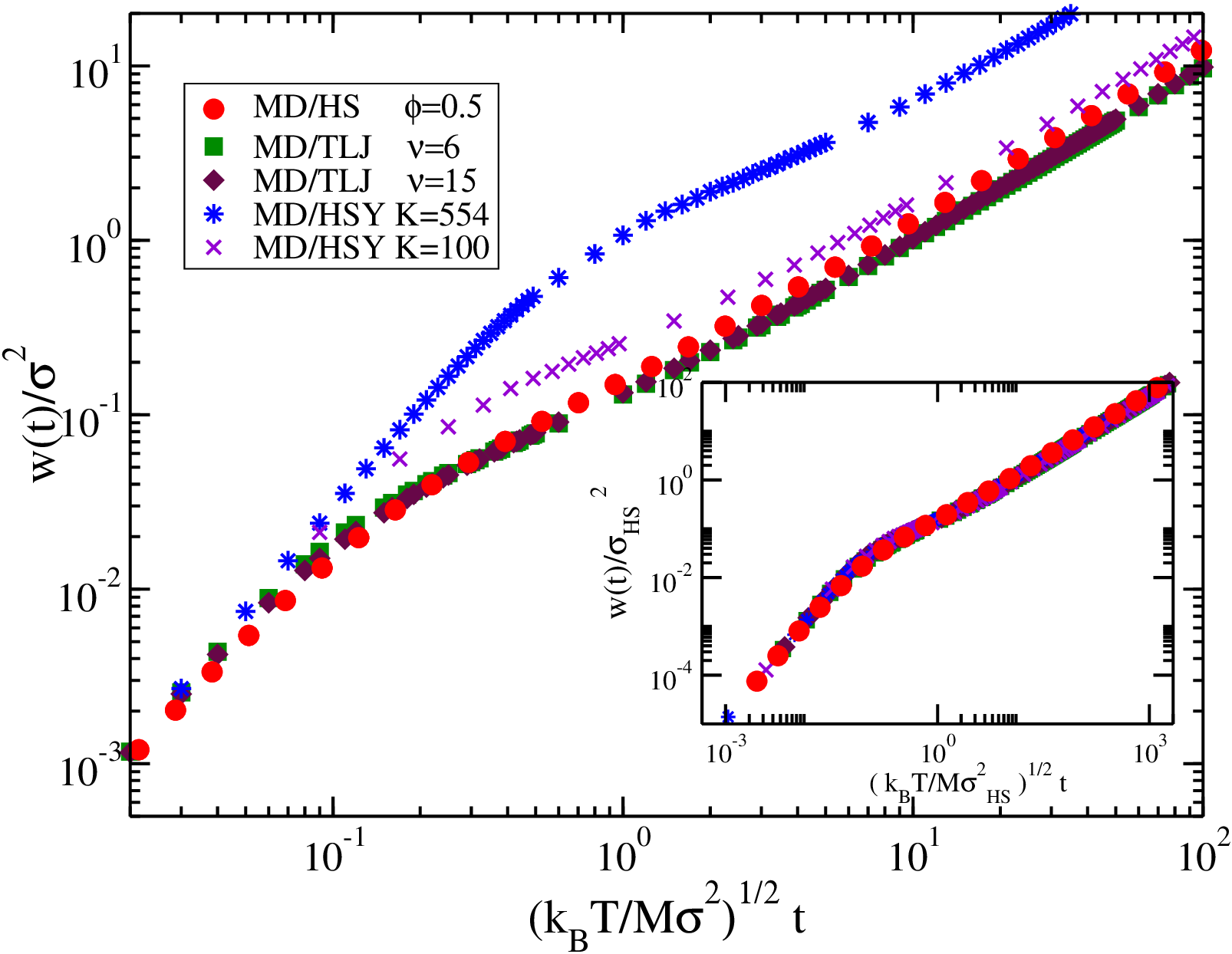}
\caption{
MSD from MD simulations
for iso-structural soft systems with equivalent $\phi_{HS}=0.49$.
The filled circles correspond to the MD simulations of a
HS fluid. The results are presented in units of $\sigma$ and
$t^{0}$. Inset: MSDs after transforming to HS units.
}
\label{wdtmd.fig}
\end{center}
\end{figure}

Thus, let us now discuss the dynamic equivalence  between
\emph{atomic} soft-sphere systems and their effective
\emph{atomic} hard-sphere counterpart. For this, let us follow the
same principle as in the Brownian case, i.e., let us express the
dynamic properties of Newtonian liquids not in terms of their
``natural'' units $\phi$, $\sigma$, and
$t^0=\sigma/v_0=\sqrt{M\sigma^2/k_BT}$, but in terms of the units
of the equivalent HS system, namely, $\phi_{HS}$, $\sigma_{HS}$,
and
$t^0_{HS}\equiv\sqrt{M\sigma^2_{HS}/k_BT}=\lambda_{\sigma}t^0$. To
see the accuracy of this predicted scaling of atomic fluids, in
Fig. \ref{wdtmd.fig} we present the MD simulation results for
the MSD of some of the illustrative model systems employed before,
namely, the TLJ system with $\nu=6$ and $10$, a HSY system with
$K=554$ and $z=0.149$,  all of
them at the effective volume fraction $\phi_{HS}=0.49$. At the
same time, we include the corresponding data for the hard spheres
fluid, obtained from event-driven MD simulations as described in
\cite{gabriel}. In the main frame of Fig. \ref{wdtmd.fig} the
MSD is scaled in the usual MD units, i.e., as $W(t)/\sigma^2$,
plotted vs. the time scaled as $t/t^0$.

As observed in the figure, the results for the various systems
clearly display the ballistic ($W(t)\sim v_0^2t^2$) and diffusive
($W(t)\sim D_Lt$) regimes characteristic of the underlying
Newtonian dynamics. They also exhibit their departure from the
exact HS results, particularly noticeable in the HSY system. In
HS units, on the other hand, the different curves in the main
panel of the figure should fall on top of the HS data, and this is
verified in the inset of the figure, which shows the MSD scaled as
$W(t)/\sigma^2_{HS} \ [=\lambda^{-2}_{\sigma} W(t)/\sigma^2]$,
plotted vs. the scaled time  as $t/t^{0}= \ [\lambda^{-1}_{\sigma}
t/t^0]$. By scaling in this manner we appreciate that all the
soft, iso-structural systems follow basically the same time
evolution, sharing, in particular, a common scaled long-time
self-diffusion coefficient.

\begin{figure}[htmb]
\centering
\includegraphics[width=7.0cm]{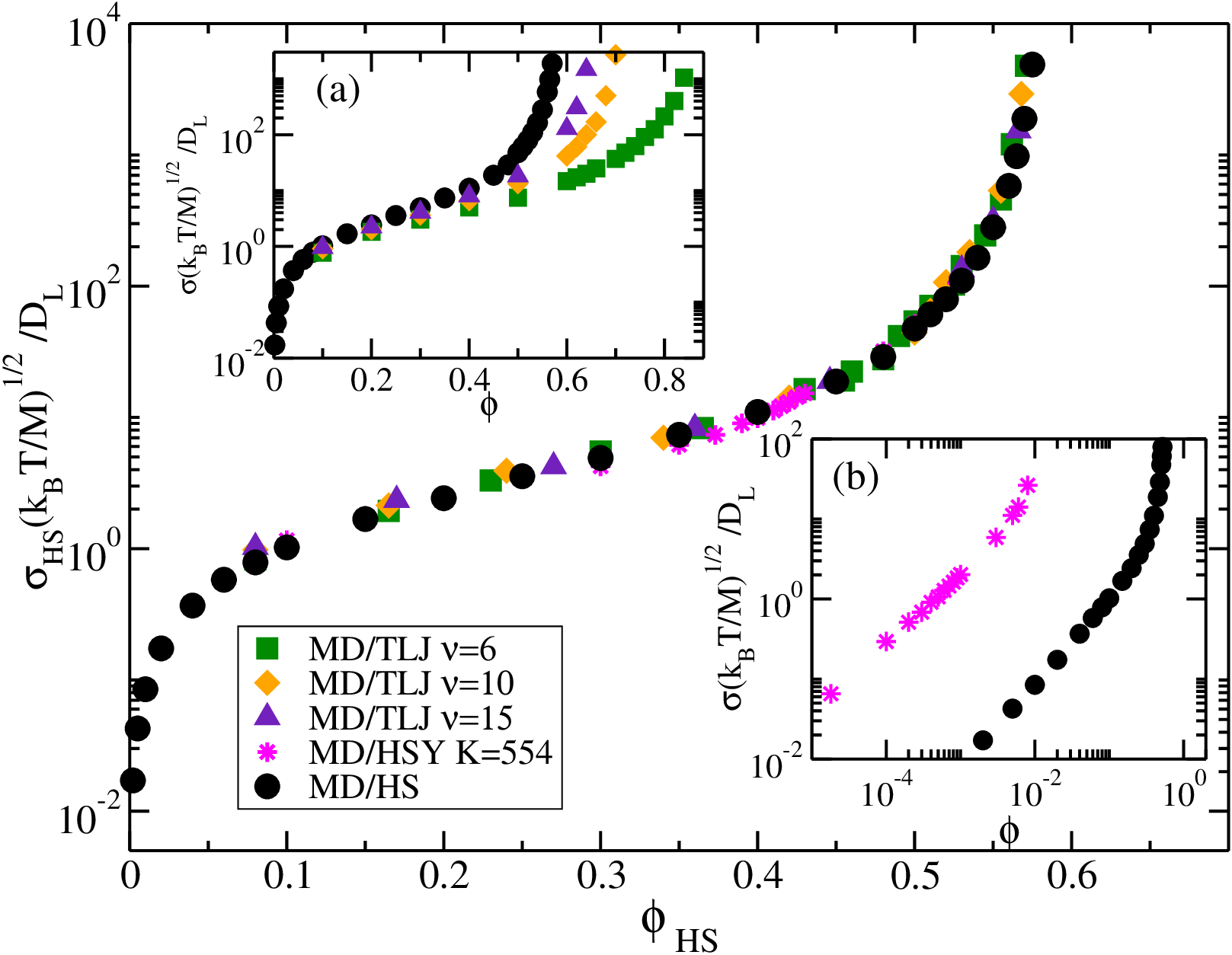}
\caption{
Inset (a): $D_{L}(\phi)$ scaled with $\sigma^{2}/t^{0}=\sigma\sqrt{k_BT/M}$ fot TLJ systems. Inset (b): Results for HSY systems. Main Panel: Results in the insets after transform to HS units. The results for strictly HS system are represented by the filled circles correspond to HS systems.
}
\label{diffmd.fig}
\end{figure}

From the long-time  limit of the results in the mainframe
of  Fig. \ref{wdtmd.fig}, one can read the value of the
long-time self-diffusion coefficient $D_L$ in units of
$\sigma^{2}/t^{0}=\sigma\sqrt{k_BT/M}$. We have collected these values of  $D_L$
for each of the systems considered here as a function of the volume fraction $\phi$ of the systems,
and the results are summarized in the two insets of Fig.
\ref{diffmd.fig}. The inset \ref{diffmd.fig} (a) contains the results for the TLJ systems,
whereas the inset \ref{diffmd.fig} (b) illustrates the noticeable contrast between the
HS and the strongly repulsive HSY system. At intermediate and
high volume fractions these data exhibit similar trends to those
observed in the corresponding BD simulations (in Fig.
\ref{diffbd.fig}(a)). The main qualitative difference between
atomic and Brownian systems can be observed at low volume
fractions, where, in contrast to Brownian systems,the atomic
$D_L$ behaves as $D_L\sim 1/\phi$ ,  at low volume fraction as the expected from the kinetic theory of dilute gases \cite{mcquarrie}.

If, on the other hand, the soft-hard dynamic equivalence were to
apply to these atomic systems, all of the data of $D_L$ displayed
in these two insets should collapse on a master curve when $D_L$
is expressed not in its ordinary atomic units, but in the
corresponding HS units, $\sigma_{HS}/t^{0}_{HS}=\sigma_{HS}\sqrt{k_BT/M}$, and plotted
not as a function of $\phi$, but of the effective  HS volume
fraction $\phi_{HS}(T^*,\phi)$. The mainframe of Fig.
\ref{diffmd.fig} plots the data in the inset in precisely this
manner. From these results we see clearly that the soft systems
follow very well the corresponding data for truly hard spheres in
all the range of volume fractions considered in the figure, thus
corroborating the expected validity of the dynamic equivalence
between soft- and hard-sphere \emph{atomic} liquids.

\begin{figure}[htmb]
\centering
\includegraphics[width=7.0cm]{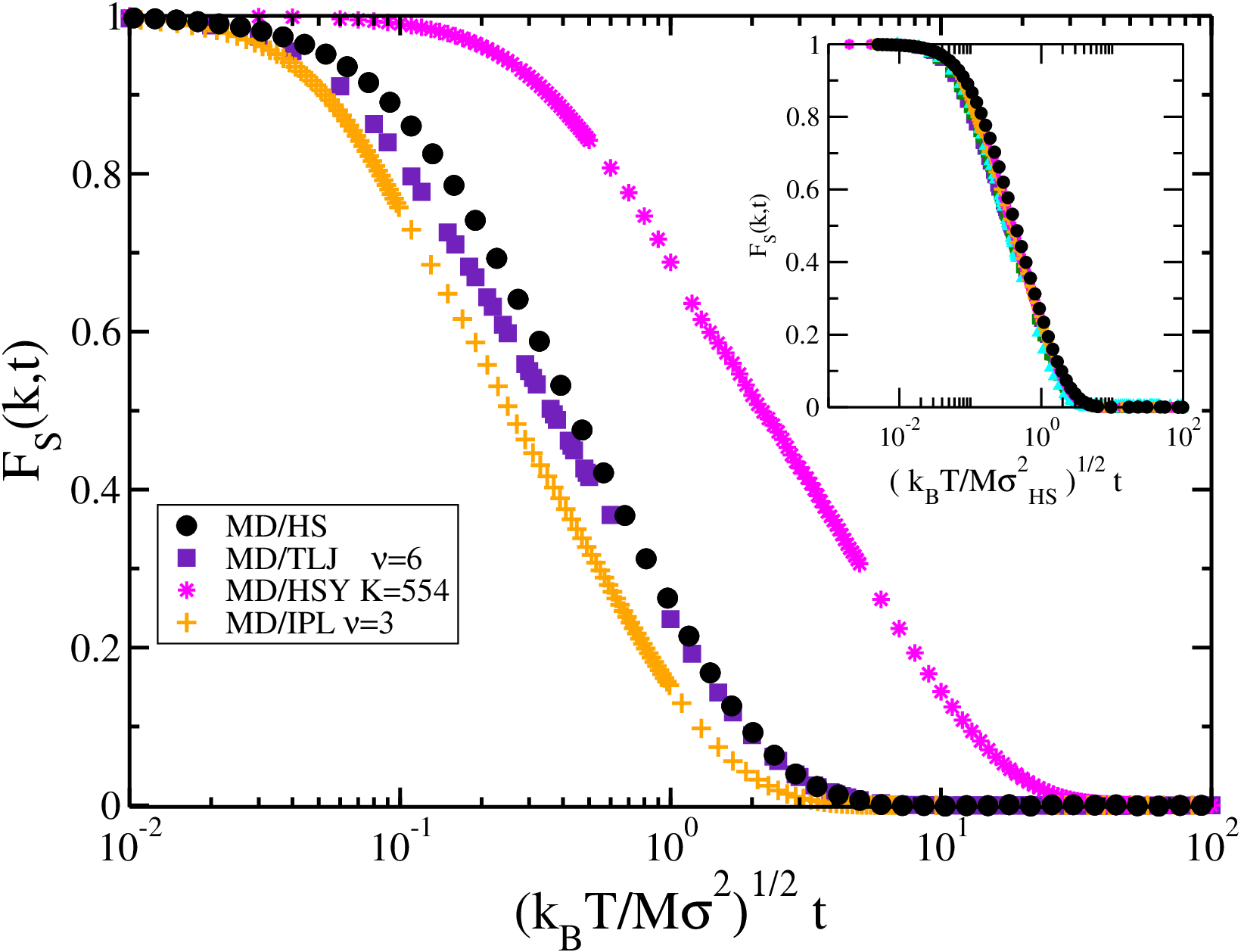}
\caption{
 $F_S(k,t)$ from
MD simulations for TLJ ($\nu=6,15$),
IPL ($\nu=3$), and HSY ($K=554$) systems
with equivalent volume fraction $\phi_{HS}=0.49$.
Time is in units of $t^{0}=\sqrt{\sigma^2M/k_BT}$.
Inset: Results after transforming to HS units via
$t_{HS}=\lambda_\sigma t^{0}$. $k=k_{max}$ is used.
}
\label{fs_md.fig}
\end{figure}

To close this section let us focus on the scaling properties of
the atomic self-ISF and its characteristic $\alpha$-relaxation
time. Thus, in figure (\ref{fs_md.fig}) we present
$F_S(k_{max},t)$ for the TLJ ($\nu=6$), IPL ($\nu=3$), HSY
($K=554$), and HS systems, all of them with equivalent HS volume
fraction $\phi_{HS}=0.49$. In the main panel, where the time is
expressed in its natural atomic units, one can see the contrast
between the various soft and the HS systems, with a scenario
rather similar to that found for Brownian fluids. In the inset, on
the other hand, we replot the same data but now as a function of
the time expressed in the HS time units
$\sqrt{M\sigma^2_{HS}/k_BT}$. There one can appreciate the
sustancial agreement between the different iso-structural systems,
which closely follow the same time-evolution in this scaled form,
with virtually the same scaled $\alpha$-relaxation time
$(k_BT/M\sigma^2_{HS})^{1/2}\tau_{\alpha}$.

\begin{figure}[htmb]
\centering
\includegraphics[width=7.0cm]{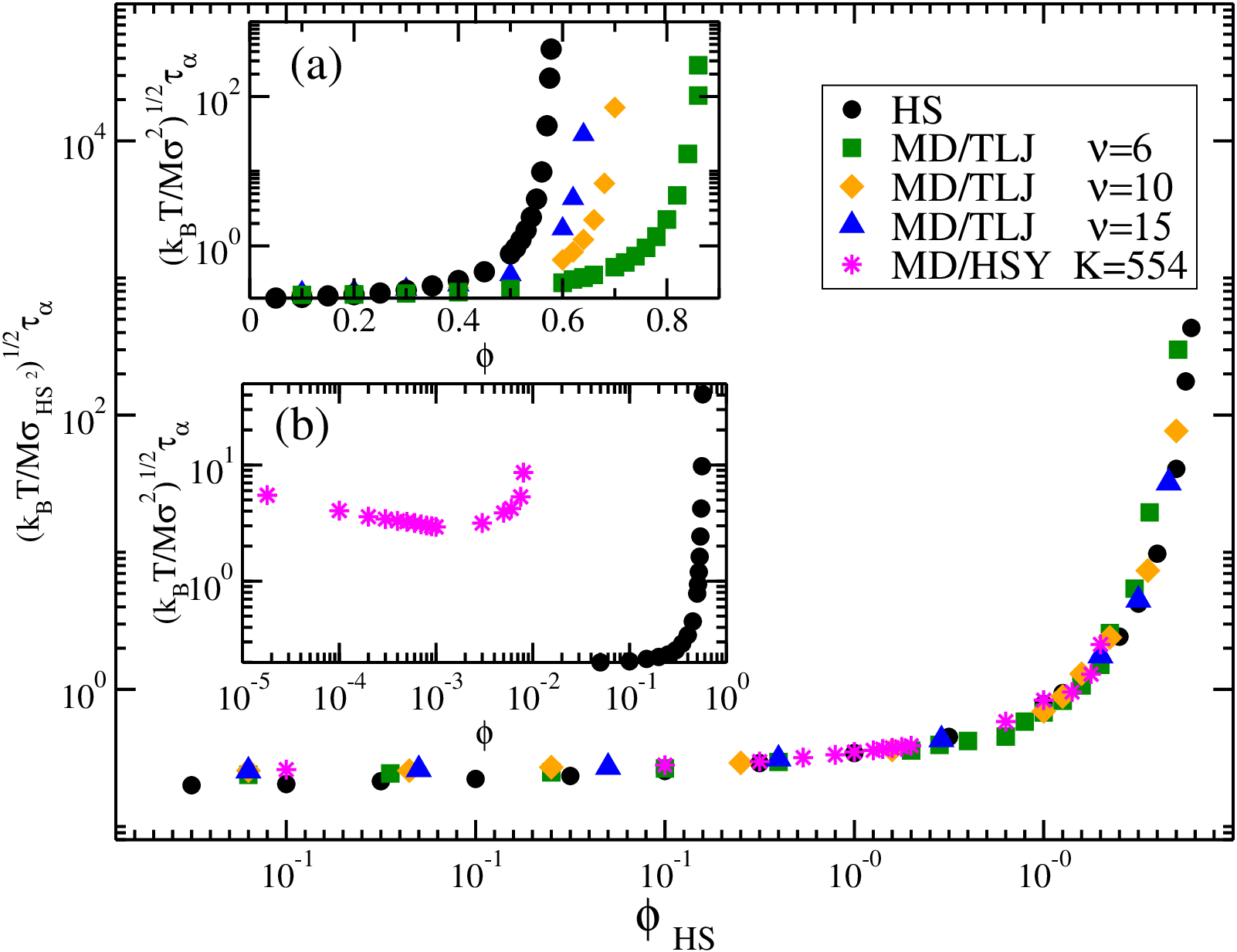}
\caption{
$\tau_{\alpha}(\phi)$, for  selected TLJ, IPL, and
HSY systems. In the three panels, the filled circles
correspond to the results for HS systems.
Inset (a): $\tau_{\alpha}(\phi)$, in MD units,
for TLJ systems with $\nu=6,10,15$.
Inset (b): $\tau_{\alpha}(\phi)$, in MD units, for
HSY systems with $K=554$ and $z=0.149$.
Main panel: Collection of results for $\tau_{\alpha}(\phi)$ in HS units.
}
\label{tau_md.fig}
\end{figure}

In Fig. \ref{tau_md.fig} we present data of $\tau_{\alpha}$
obtained from similar simulations of $F_S(k_{max},t)$ for the
various systems carried out varying the density. These data are
plotted as a function of the respective volume fraction $\phi$ in
the ordinary atomic units (insets), and normalized with the HS
units (main panel).  In general, again, the trends here resemble
closely those of the Brownian systems previously discussed, in the
sense that the completely disperse unscaled data in the insets, in the main
panel, collapse nicely onto a master curve that coincides, of
course, with the MD data of strictly hard spheres (black symbols).
This concludes the present discussion of the soft-hard dynamic
equivalence in both, atomic and Brownian liquids. In what follows
we discuss a related but fundamentally different scaling.

\section{Brownian-Atomic Dynamic Equivalence}\label{sec5}

As mentioned in the introduction, the second fundamental challenge
in understanding the relationship between dynamic arrest phenomena
in colloidal systems and the glass transition in atomic liquids is
to determine the role played by the underlying (Brownian vs.
Newtonian) microscopic dynamics. The results presented in the
previous section provide one step forward in this direction, since
they demonstrate that the criterion to unify a rather diverse set
of soft-sphere systems in a so-called dynamic universality class
is actually independent of the Brownian or Newtonian nature of
their microscopic dynamics. As a result, for example, the data for
the long-time self-diffusion coefficient $D_L$ of various Brownian
systems, displayed in Fig. \ref{diffbd.fig}(a), collapse onto the
master curve of Fig. \ref{diffbd.fig}(b). In its turn, the data
for $D_L$ of the \emph{atomic version} of the same systems, shown
in the insets of Fig. \ref{diffmd.fig}, collapse onto their own
master curve, displayed in the main frame of the same figure.

An important question, however, is left open by these results. It
refers to the possibility that a fundamental relationship can be
established, now between those two master curves (in Figs.
\ref{diffbd.fig}(b) and \ref{diffmd.fig} respectively), which
would unify the dynamics of an atomic liquid with the dynamics of
its Brownian counterpart in an unambiguous and precise manner.
This question was largely answered theoretically in recent
attempts of our group to extend the SCGLE theory of colloid
dynamics to atomic systems \cite{atomic1,atomic2}. There it was
established that such a relationship is provided by the
recognition that the (density- and temperature-dependent)
self-diffusion coefficient of an \emph{atomic} liquid, determined
by kinetic theoretical arguments as

\begin{equation}
D^0=\frac{3}{8\sqrt{\pi}} \left(\frac{k_BT}{M}\right)^{1/2}
\left(\frac{1}{n\sigma_{HS}^2}\right),
\label{d0kin.eq}
\end{equation}
plays the role of the short-time self-diffusion coefficient $D^0$
in Brownian systems.

As a result, in Refs.  \cite{atomic1,atomic2} it was predicted,
for example, that the ratio $D^*\equiv D_L/ D^0$ of the long-time
to the short-time self-diffusion coefficients of a Brownian system
must be identical to the long-time self-diffusion coefficient of
the corresponding atomic liquid, scaled with this kinetic
theoretical value of $D^0$. Testing this particular prediction is,
of course, very straightforward, and can be done by normalizing
the data of $D_L$ for the atomic systems in the insets of Fig.
\ref{diffmd.fig}, with the value of $D_0$ given by Eq.
(\ref{d0kin.eq}). The resulting scaled data should then coincide
with the corresponding results for $D^*$ of the Brownian version
of the same systems in Fig. \ref{diffbd.fig}(a).

\begin{figure}[htmb]
\centering
\includegraphics[width=7.0cm]{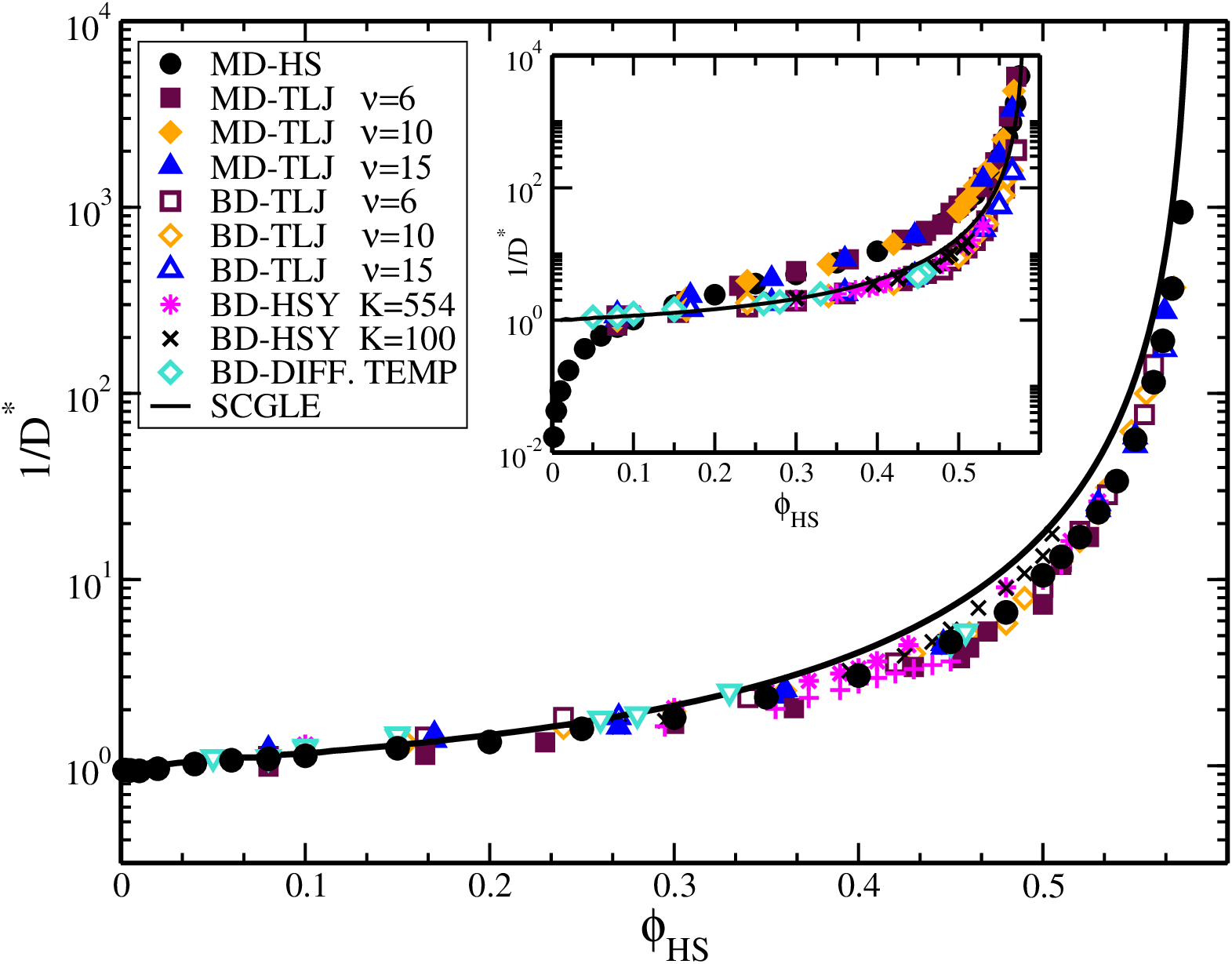}
\caption{ Scaled results for the long time diffusion coefficient for colloid and atomic
liquids. In the main panel $D^{*}(\phi_{HS})=D_{L}/D^{0}$ is plotted using the corresponding definition of $D^{0}$. Inset: $D_{L}(\phi_{HS})$ in HS units as reported in the previous Figs. 8(b) and 13.
 } \label{dl_bd_md.fig}
\end{figure}

The same comparison, however, can be done more directly if we take
the same data, but after they have been collapsed onto their
respective master curve in Figs. \ref{diffbd.fig}(b) and
\ref{diffmd.fig}. Thus, in the inset of Fig. \ref{dl_bd_md.fig} we
reproduce these two master curves, to highlight the different
behavior of atomic and Brownian liquids regarding the density
dependence of the data for $D_L$ expressed in the effective HS
units. The next step is then to scale the atomic data for $D_L$ as
$D^*\equiv D_L/ D^0$ with $D_0$ given by Eq. (\ref{d0kin.eq}). The
result of this scaling is that the original atomic master curve
now coincides with the original Brownian master curve, as
demonstrated in the main frame of Fig. \ref{dl_bd_md.fig}. Since
the Brownian data for $D_L$ were already expressed as $D^*\equiv
D_L/ D^0$, this collapse between both master curves illustrates
the accuracy of the predicted dynamic equivalence between atomic
and Brownian liquids.

\begin{figure}[htmb]
\centering
\includegraphics[width=7.0cm]{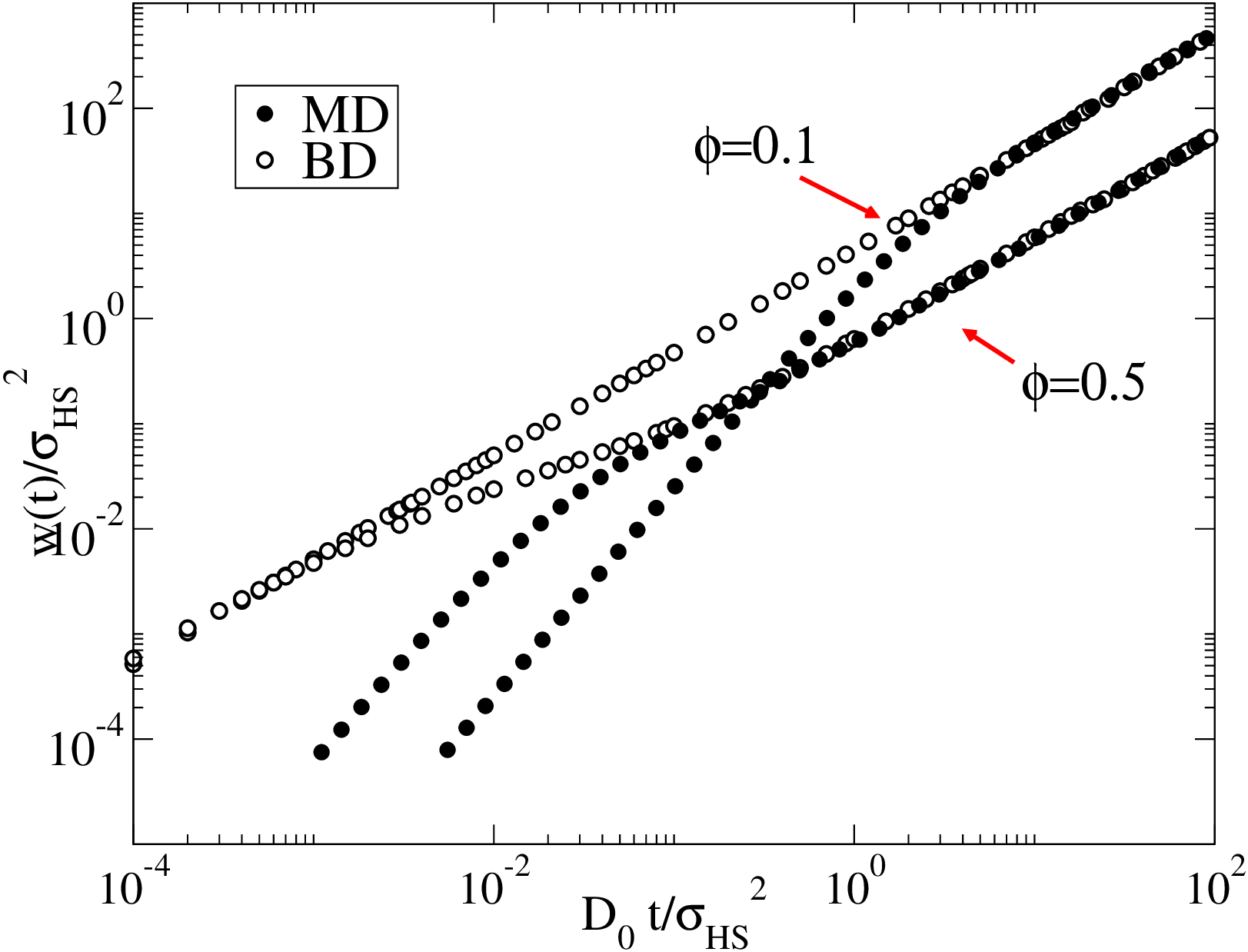}
\caption{
Mean squared displacement, from
molecular dynamics (solid symbols) and Brownian dynamics (
empty symbols)simulation,  for equivalent hard-sphere systems, for states
corresponding to $\phi_{HS}= 0.1$ and $0.5$. }
\label{wt_bd_md.fig}
\end{figure}

As originally proposed, however, the atomic-Brownian scaling
extends to the time-dependent dynamic properties, such as the mean
squared displacement and the self intermediate scattering
functions. Thus, in Refs. \cite{atomic1,atomic2}
it was predicted that these properties of a given atomic liquid,
at times $t$ much longer than the mean free time, and with $t$
scaled with $D^0$ in Eq. (\ref{d0kin.eq}), will be
indistinguishable from those of its Brownian counterpart. To
illustrate such condition, the atomic and  Brownian MSD of the TLJ potential with $\nu=6$ and effective HS volume fractions $\phi_{HS} = 0.1$ and $0.5$, are
presented in Fig. \ref{wt_bd_md.fig} in the format
$W(t)/\sigma^2_{HS}$ vs. $D_0t/\sigma_{HS}^2$ (i.e. in HS units). The results
clearly show that despite the differences at short times
(ballistic vs. diffusive),  the MSDs of iso-structural systems do
collapse onto each other in the long-time regime.

\begin{figure}[htmb]
\centering
\includegraphics[width=7.0cm]{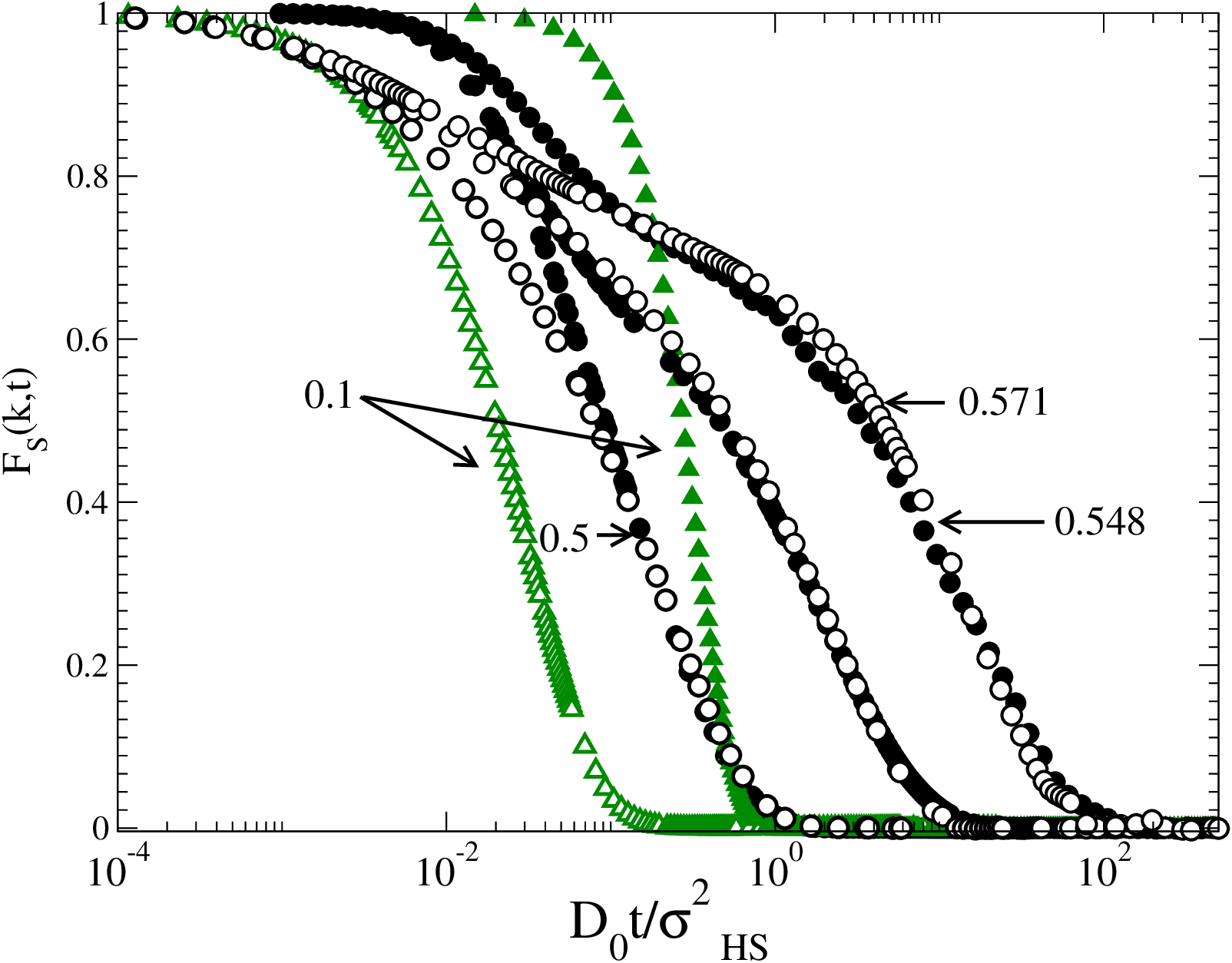}
\caption{
MD (solid symbols) and BD
(empty symbols) simulation results for the
$F_S(k,t)$ of equivalent HS liquids at
volume fractions  $\phi_{HS} = 0.10, 0.50, 0.548$ and $0.571$,
evaluated  at the main peak of the static structure factor and plotted
as a function of the dimensionless time $D^0t/\sigma^2_{HS}$.
}
\label{fself_collaps.fig}
\end{figure}

This long-time atomic-Brownian scaling can also be observed in
$F_S(k,t)$. However, in contrast with the MSD, in which this
scaling holds at all effective volume fractions, in the case of
$F_S(k,t)$ this scaling only holds above a certain threshold
effective volume fractions, corresponding essentially to the
metastable liquid regime. This is illustrated in Fig.
\ref{fself_collaps.fig}, where we present  molecular and
Brownian dynamics results for this property at three volume fractions of the effective hard-sphere
liquid, $\phi_{HS}= 0.50, 0.548$, and
$0.571$ (generated, in reality, with the dynamically equivalent
TLJ potential with $\nu=6$). In this figure $F_S(k,t)$ is
plotted as a function of the dimensionless time $D^0
t/\sigma^2_{HS}$, with the corresponding $D^0$ for each dynamics
(i.e., given by Eq. (\ref{d0kin.eq}) in the atomic version of the
system).

As indicated above, this long-time dynamic equivalence between
atomic and Brownian liquids is not observed in $F_S(k,t)$ at low
volume fractions corresponding to the stable fluid regime (i.e.,
for $\phi_{HS}\lesssim 0.5$). This is illustrated in Fig.
\ref{fself_collaps.fig} with the atomic and Brownian results for
$F_S(k,t)$ corresponding to $\phi_{HS}= 0.1$, which totally
fail to collapse on top of each other, especially at long times. The reason for this
deviation from the long-time dynamic equivalence at low volume
fractions is that in this regime, the decay of the atomic
$F_S(k,t)$ to a value $\approx e^{-1}$ occurs within times
comparable to the mean free time $\tau_0$ and is, hence,
intrinsically ballistic. It is only at higher volume fractions
that this long-time dynamic equivalence is fully exhibited by the
diffusive decay of $F_S(k,t)$, as illustrated by the three largest
volume fractions in the figure.

\begin{figure}[htmb]
\centering
\includegraphics[width=7.0cm]{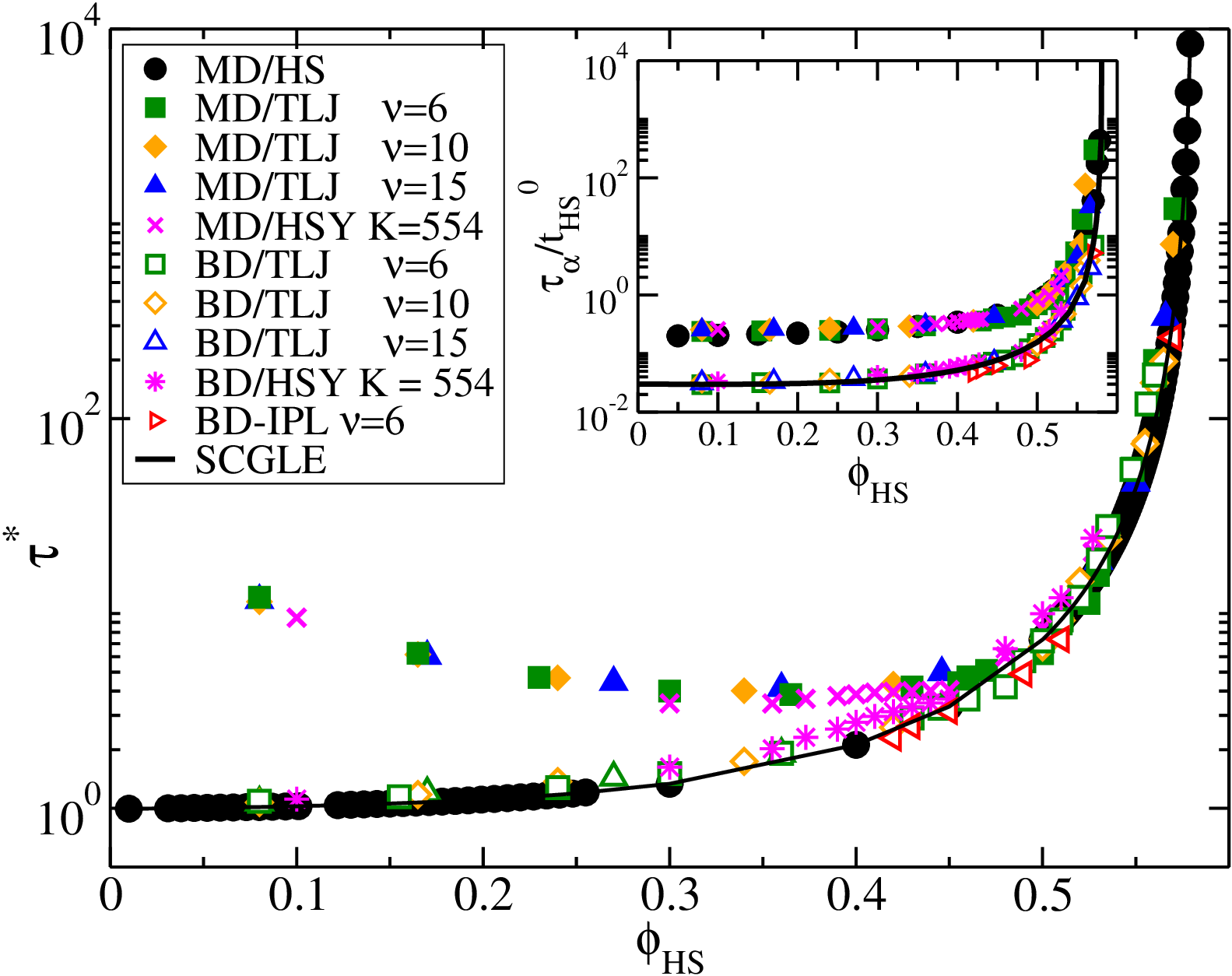}
\caption{ Inset: $\alpha$-relaxation time $\tau_{\alpha}(\phi_{HS})$
for Brownian and Molecular liquids in HS units as reported in the previous  Figs. 11 and 15. Main panel: Data presented as $\tau^*\equiv k^2D_0\tau_{\alpha}$.
The solid curve is the prediction for $\tau^{*}$ for the colloidal
SCGLE.
 }
\label{tau_bd_md.fig}
\end{figure}

The observations above can also be summarized by comparing the
volume fraction dependence of the relaxation time $\tau_{\alpha}$
of the molecular and Brownian versions of the various soft-sphere
systems discussed in the previous section. Such results were
summarized in the Brownian and atomic master curves presented,
respectively, in Figs. (11) and (15), which we now put together in
the inset of Fig. \ref{tau_bd_md.fig}. To exhibit the long-time
dynamic equivalence between atomic and Brownian fluids, the same
results are presented again in the main panel of the figure, but
now scaled as  $\tau^*\equiv k^2 D^0 \tau_{\alpha}$ with $k=k_{max}$ in HS units. From the
corresponding comparison one can see that this long-time dynamic
equivalence, which manifests itself in the collapse of the
molecular and Brownian dynamics data for $\tau^*$, holds only above a threshold volume fraction, roughly
located near the freezing transition of the HS liquid
($\phi_{HS}\approx0.5$). The qualitative difference between atomic
and Brownian systems observed in the results for $\tau^*$ for
volume fractions below this threshold are explained in the
different low-density limit of  $\tau^*$ in each case. For a
Brownian liquid $\tau_{\alpha} \rightarrow 1/k^2 D^0$ as
$\phi\rightarrow 0$, with a $\phi$-independent short-time
diffusion coefficient $D^0$, so that $\tau^* \rightarrow 1$ as
$\phi\rightarrow 0$. For atomic liquids, however, $\tau^*
\rightarrow \sqrt{2\pi}k\sigma/16\phi$ in the same limit, where we
have taken into account the fact that in this case the short-time
diffusion coefficient $D^0$ is given by the kinetic-theoretical
result in equation (\ref{d0kin.eq}).

\vskip0.5cm

\vskip1cm

\section{Summary}\label{sec6}

In this work we have presented an extended simulation study on the
static and dynamic equivalence between model fluids with soft
repulsive potentials of varying range and strength, governed by
either Brownian or atomic microscopic dynamics. The dynamic
equivalence investigated here relies heavily on the concept of
static structural equivalence proposed by Weeks, Chandler and
Andersen, i.e., on the notion that the thermodynamic and
structural properties of fluids formed by moderately  soft
particles can be expressed in terms of the properties of the
hard-sphere liquid.

Thus, the present work started by reviewing such concept of
structural mapping, explaining how this idea is extended to map
the static structure of, in principle, {\em any} soft-sphere
liquid onto the properties of an effective hard-sphere liquid. We
provided details of the method adopted to identify structurally
equivalent systems, which works better than other traditional
approaches, such as the blip function method, especially for
long-ranged potentials. We also explained how to extend this
mapping procedure to the case of polydisperse systems, which
allowed us to study highly concentrated systems beyond freezing,
i.e., in the metastable liquid fase.

This structural equivalence was then employed in the study of the
dynamic equivalence between iso-structural {\em colloidal} fluids.
For this, simple rules are provided to map the units of length and
time for soft systems to those of the equivalent HS systems. Our
extensive Brownian dynamics simulations verified, for example,
that the mean square displacement and the self intermediate
scattering functions of iso-structural systems follow the same
time evolution when they are expressed in the units of the
equivalent HS system. In particular, data were provided as a
function of the volume fraction, covering from dilute to highly
concentrated conditions, to show that long time properties such as
the long time self diffusion coefficient and the
$\alpha$-relaxation time of soft colloidal systems collapse onto a
master curve when plotted as a function of the (density- and
temperature-dependent) effective HS volume fraction. Furthermore,
we showed that this dynamic equivalence extends over to the domain
of {\em atomic} fluids. Thus, in Sect. \ref{sec4} we demonstrated
that the criterion that unifies the soft-sphere systems in the
hard-sphere dynamic universality class is actually independent of
the Brownian or Newtonian nature of their microscopic dynamics, in
the sense that the properties of Brownian systems collapse onto a
given master curve, as in Fig. \ref{diffbd.fig}(b), and the same
applies to the \emph{atomic version} of the same systems, whose
properties collapse onto their own master curve, as in Fig.
\ref{diffmd.fig}.

The findings just described then reveal that the dynamics of soft
systems can be mapped onto those of HS systems, regardless of the
underlying Brownian or Newtonian dynamics, i.e., both type of
systems satisfy a soft-hard dynamics equivalence. Going further,
however, the inspection of the long-time behavior of atomic
liquids revealed another connection, this time between colloidal
and atomic fluids. This connection was established  once the
atomic-liquid analog of the Brownian short-time self-diffusion
coefficient $D^0$ is identified with the the value predicted by
the kinetic theory of gases, i.e., by $D^0$ given in Eq.
(\ref{d0kin.eq}). The simulation evidence that we provide here
corroborate that at least for model liquids whose structure is
dominated by (soft- or hard-sphere) repulsive interactions, the
long-time dynamics of an atomic liquid is indistinguishable from
the dynamics of the colloidal system with the same inter-particle
interactions. As a consequence, some dimensionless long-time
dynamic properties, such as $D^*=D_L/D^0$ (with the proper
identification of $D^0$) and $\tau^*(k)=k^2D^0\tau_{\alpha}$ (only
in the supercooled liquid regime) will exhibit, just like the
equilibrium thermodynamic and structural properties, the same
independence from the short-time microscopic dynamics, which
otherwise distinguishes atomic from colloidal systems.

Let us clarify that for the present study we chose to focus on a
well-defined set of model systems, namely, systems with purely
repulsive soft-sphere interactions of arbitrary range. This
excludes from this study the consideration, for example, of the
possibility of full overlap between particles, characteristic of
ultrasoft interactions \cite{likos}. Hence, a pending question is
the degree at which the scalings discussed in this work will apply
to these systems. Similarly, the absence of attractive forces in
our working examples leaves open the issue of the possible
extension of these scalings to systems that involve attractive
interparticle forces. We have performed, however, preliminary
calculations with both, ultrasoft systems and systems with
attractive interactions, and the results suggest that these
scalings have a much wider range of applicability. Another
important aspect that requires further discussion is the
relationship of the dynamic scalings discussed in this work and
other scalings discussed in the literature. We have mostly in mind
the interesting concept of strongly correlated liquids, developed
by J. Dyre and collaborators
\cite{dyrejpcm2013,gnan1,gnan2,dyreprx}. It will be interesting to
clarify, for example, if the concept of isomorphic states defined
from their perspective, is related to our findings that
isostructurality is also associated with isodiffusivity.

On the other hand, and as a final remark, let us mention that here
we referred to Brownian liquids as colloidal suspensions, with the
intention to connect with real physical systems. In reality,
however, we really meant colloidal systems for which hydrodynamic
interactions can be neglected, since these important effects were
completely ignored in our Brownian dynamics simulations. We
expect, however, that most of our conclusions will apply to real
colloidal systems in which hydrodynamic interactions are
important, as long as we identify the parameter $D^0$ not with the
value of the long-time self-diffusion coefficient $D_L$ at
infinite dilution ($\phi=0$), but with the $\phi$-dependent
short-time self-diffusion coefficient $D_S(\phi)$ \cite{prlhi},
which under some circumstances can be independently determined
either theoretical or experimental methods. Fig. 3 of Ref.
\cite{gabriel} illustrates the effectiveness of this hydrodynamic
scaling, which is expected to expand the range of application of
the soft-to-hard and Brownian-to-atomic dynamic equivalences
discussed in this work.

\bigskip
\bigskip
\bigskip

ACKNOWLEDGMENTS: This work was supported by the Consejo Nacional de
Ciencia y Tecnolog\'{\i}a (CONACYT, M\'{e}xico), through grants No.
132540 and 182132.


\vskip1cm


\begin{thebibliography}{99}

\bibitem{angellreview1}
Angell C. A., Ngai K. L., McKenna G. B.,
McMillan P. F. and Martin S. F.,
J. Appl. Phys. \textbf{88} 3113 (2000).

\bibitem{edigerreview1}
M. D. Ediger,  C. A. Angell, and S. R. Nagel,
J. Phys. Chem.\textbf{100}, 13200 (1996)

\bibitem{ngaireview1}
K. L. Ngai, D. Prevosto, S. Capaccioli and C. M. Roland,
J. Phys.: Condens. Matter \textbf{20}, 244125 (2008)

\bibitem{angell}
C. A. Angell,
Science {\bf 267}, 1924 (1995).

\bibitem{debenedetti}
P. G. Debenedetti and F. H. Stillinger,
Nature {\bf 410}, 359 (2001).

\bibitem{sciortinotartaglia}
F. Sciortino and P. Tartaglia,
Adv. Phys. {\bf 54}, 471 (2005).

\bibitem{jammingrheology}
\emph{Jamming and Rheology: Constrained Dynamics on
Microscopic and Macroscopic Scales},
edited by A. J. Liu and S. R. Nagel
(Taylor \& Francis, New York, 2001).

\bibitem{xu}
N. Xu, T. K. Haxton, A. J. Liu, and S. R. Nagel,
Phys. Rev. Lett. \textbf{103} 245701 (2009).

\bibitem{berthierwitten1}
L. Berthier and T. A. Witten,
Europhys. Lett. {\bf 86}, 10001 (2009).

\bibitem{lowenhansenroux}
H. L\"owen, J. P. Hansen, and J. N. Roux,
Phys. Rev. A \textbf{44}, 1169 (1991).

\bibitem{szamelflenner}
G. Szamel and E. Flenner,
Europhys. Lett., \textbf{67}, 779 (2004).

\bibitem{puertasaging}
A. M. Puertas,
J. Phys.: Condens. Matter \textbf{22}, 104121 (2010).

\bibitem{goetze1}
W. G\"{o}tze,
in {\em Liquids, Freezing and Glass Transition},
edited by J. P. Hansen, D. Levesque, and J. Zinn-Justin
(North-Holland, Amsterdam, 1991).

\bibitem{szamellowen}
G. Szamel and H. L\"owen,
Phys. Rev. A \textbf{44}, 8215 (1991).

\bibitem{dyrejpcm2013}
L. B{\o}hling et al., J. Phys.: Condens. Matter, \textbf{25},
032101 (2013).

\bibitem{boonyip}
J. L. Boon and S. Yip,
{\em Molecular Hydrodynamics}{\it \ }
(Dover Publications Inc. N. Y., 1980).

\bibitem{faraday}
M. Medina-Noyola,
Faraday Discuss. Chem. Soc. {\bf 83}, 21 (1987).

\bibitem{delrio}
M. Medina-Noyola and J. L. del R\'{i}o-Correa,
{\em Physica 146A}, 483 (1987).

\bibitem{scgle0}
L. Yeomans-Reyna and M. Medina-Noyola,
Phys. Rev. E {\bf 62}, 3382 (2000).

\bibitem{scgle1}
L. Yeomans-Reyna and M. Medina-Noyola,
Phys. Rev. E {\bf 64}, 066114 (2001).

\bibitem{scgle2}
L. Yeomans-Reyna, H. Acu\~{n}a-Campa, F. Guevara-Rodr\'{\i}guez,
and M. Medina-Noyola,
Phys. Rev. E {\bf 67}, 021108 (2003).

\bibitem{rmf}
P.E. Ram\'{\i}rez-Gonz\'alez {\it et al.},
Rev. Mex. F\'{\i}sica \textbf{53}, 327  (2007).

\bibitem{todos1}
L. Yeomans-Reyna {\it et al.},
Phys. Rev. E \textbf{76}, 041504 (2007).

\bibitem{todos2} R. Ju\'arez-Maldonado {\it et al.}, Phys. Rev. E {\bf 76},
062502 (2007).

\bibitem{noneqscgle0} P. E. Ram\'irez-Gonz\'alez and M. Medina-Noyola,
Phys. Rev. E \textbf{82}, 061503 (2010); ibid. Phys. Rev. E \textbf{82}, 061504
(2010).

\bibitem{noneqscgle1} P. E. Ram\'irez-Gonz\'alez and M. Medina-Noyola,
Phys. Rev. E \textbf{82}, 061503 (2010); ibid. Phys. Rev. E \textbf{82}, 061504
(2010).

\bibitem{soft1} P. E. Ram\'irez-Gonz\'alez  and M. Medina-Noyola, J. Phys.:
Cond. Matter, {\bf 21}, 75101 (2009).

\bibitem{soft2} P. E. Ram\'irez-Gonz\'alez, L. L\'opez-Flores, H. Acu\~na-Campa,
and M. Medina-Noyola,  Phys. Rev. Lett. \textbf{107}, 155701 (2011).

\bibitem{awc} H. C.Andersen, J. D. Weeks and D. Chandler, Phys. Rev. A
\textbf{4}, 1597 (1971).

\bibitem{hansen} J.P. Hansen and I.R. McDonald, {\em \ Theory of Simple Liquids}
(Academic
Press Inc., 1976)

\bibitem{atomic1} P. Mendoza-M\'endez, L. L\'opez-Flores, A. Vizcarra-Rend\'on,
L. E. S\'anchez-D\'iaz, and M. Medina-Noyola, arXiv:1203.3893v1 [cond-mat.soft].

\bibitem{atomic2} L. L\'opez-Flores, L. L. Yeomans-Reyna, M.
Ch\'avez-P\'aez, and M. Medina-Noyola, J. Phys.: Condens. Matter,
{\bf 24} 375107 (2012).

\bibitem{nagele0}  G. N\"{a}gele, {\em Phys. Rep. }{\bf 272}, 215 (1996).

\bibitem{gaylor} K.J. Gaylor, I.K. Snook, W. van Megen, and R.O. Watts, J.
Phys. A \textbf{13}, 2513 (1980).

\bibitem{allen} M.P. Allen and D.J. Tildesley,
{\it Computer Simulation of Liquids}, (Oxford, University Press, 1887.)

\bibitem{swapmc}
T.S. Grigera and G. Parisi,
Phys. Rev. E {\bf 63}, 45102(R), (2001).

\bibitem{wdtblocks}
D. Dubbeldam, D.C. Ford, D.E. Ellis, and R.Q. Snurr,
Mol. Phys. {\bf 35}, 1084 (2009).

\bibitem{qls}
J.S. van Duijneveldt and D. Frenkel,
J. Chem. Phys. {\bf 96}, 4655 (1992).

\bibitem{gabriel} G. P\'erez-\'Angel et al.,
Phys. Rev. E {\bf 83}, 060501 (2011).

\bibitem{percusyevick}  J. K. Percus and G. J. Yevick,
Phys. Rev. {\bf 110}, 1 (1957).

\bibitem{wertheim} M. S. Wertheim, Phys. Rev. Lett.  {\bf 10}, 321 (1963).

\bibitem{verletweis} L. Verlet and J.-J. Weis, Phys. Rev. A  {\bf  5} 939 (1972).
\bibitem{dynamicequivalence}  F. de J. Guevara-Rodr\'{\i}guez and M.
Medina-Noyola,  Phys. Rev. E {\bf   68}, 011405 (2003).

\bibitem{lebowitzhs2mixt} J. L. Lebowitz, Phys. Rev. \textbf{133}, A825 (1972).

\bibitem{hiroike} K. Hiroike, Mol. Phys. {\bf 33}, 1519 (1977); ibid, J. Phys. Soc. Jpn.
{\bf 27}, 1415 (1969).

\bibitem{williamsvanmegen} S. R. Williams and W. van Megen, Phys. Rev. E \textbf{64},
041502 (2001).

\bibitem{pusey0}  P. N.  Pusey in {\em Liquids, Freezing and Glass Transition}, edited by
J. P. Hansen, D. Levesque, and J. Zinn-Justin (Elsevier, Amsterdam,
1991), Chap. 10.

\bibitem{mcquarrie} D.A. McQuarrie, \textsl{Statistical Mechanics}, Harper and Row,
N.Y. (1975).

\bibitem{atomicbrownian}
L. L\'opez-Flores, et al., Europhys. Letts. {\bf 99}, 46001
(2012).

\bibitem{likos} C. N. Likos, Soft Matter, \textbf{2}, 478 (2006).

\bibitem{gnan1} N. Gnan et al.,  J. Chem. Phys. \textbf{131}, 234504 (2009).

\bibitem{gnan2} N. Gnan et al., Phys. Rev. Lett. \textbf{104},
125902 (2010).

\bibitem{dyreprx} T. S. Ingebrigtsen, T. B. Schr{\o}der, and J. C. Dyre, Phys. Rev. X
\textbf{2}, 011011 (2013)



\bibitem{prlhi} M. Medina-Noyola, Phys. Rev. Lett. \textbf{60}, 2705 (1988).


\end{thebibliography}
\end{document}